\documentclass[11pt,a4paper]{article}
\usepackage[all,cmtip]{xy}
\usepackage{amssymb}
\usepackage{amscd}
\usepackage{pdfsync}
\begin{document}

%
%
%
%


\newcommand{\Si}{\Sigma}
\newcommand{\tr}{{\rm tr}}
\newcommand{\ad}{{\rm ad}}
\newcommand{\Ad}{{\rm Ad}}
\newcommand{\ti}[1]{\tilde{#1}}
\newcommand{\om}{\omega}
\newcommand{\Om}{\Omega}
\newcommand{\de}{\delta}
\newcommand{\al}{\alpha}
\newcommand{\te}{\theta}
\newcommand{\vth}{\vartheta}
\newcommand{\be}{\beta}
\newcommand{\la}{\lambda}
\newcommand{\La}{\Lambda}
\newcommand{\D}{\Delta}
\newcommand{\ve}{\varepsilon}
\newcommand{\ep}{\epsilon}
\newcommand{\vf}{\varphi}
\newcommand{\G}{\Gamma}
\newcommand{\ka}{\kappa}
\newcommand{\ip}{\hat{\upsilon}}
\newcommand{\Ip}{\hat{\Upsilon}}
\newcommand{\ga}{\gamma}
\newcommand{\ze}{\zeta}
\newcommand{\si}{\sigma}
\newcommand{\wt}[1]{\widetilde{#1}}
\newcommand{\brc}[1]{\left(#1\right)}
\newcommand{\rme}{\textrm{e}}
\newcommand{\rmd}{\textrm{d}}
\newcommand{\bfi}[1]{\left\{ #1\right\}}
\newcommand{\bsq}[1]{\left[#1\right]}
\newcommand{\matr}[2]{\begin{array}{#1}#2\end{array}}
\newcommand{\svs}{\\[7pt]}
\def\bfa{{\bf a}}
\def\bfb{{\bf b}}
\def\bfc{{\bf c}}
\def\bfd{{\bf d}}
\def\bfe{{\bf e}}
\def\bff{{\bf f}}
\def\bfm{{\bf m}}
\def\bfn{{\bf n}}
\def\bfp{{\bf p}}
\def\bfq{{\bf q}}
\def\bfu{{\bf u}}
\def\bfv{{\bf v}}
\def\bft{{\bf t}}
\def\bfx{{\bf x}}
\def\bfy{{\bf y}}
\def\bfg{{\bf g}}
\def\bfS{{\bf S}}
\def\bfJ{{\bf J}}
\def\bfC{{\bf C}}
\def\bfF{{\bf F}}
\def\bfE{{\bf E}}
\def\bv{{\bf v}}
\def\bfW{{\bf W}}
\def\bfphi{{\bf \phi}}
\def\hP{\hat{P}}
\def\hS{\hat{S}}
\def\bhS{\bf\hat{S}}

\def\cA{{\cal A}}
\def\cB{{\cal B}}
\def\cC{{\cal C}}
\def\cD{{\cal D}}
\def\cF{{\cal F}}
\def\cE{{\cal E}}
\def\cH{{\cal H}}
\def\cJ{{\cal J}}
\def\cI{{\cal I}}
\def\cO{{\cal O}}
\def\cS{{\cal S}}
\def\cG{{\cal G}}
\def\cL{{\cal L}}
\def\cM{{\cal M}}
\def\cP{{\cal P}}
\def\cQ{{\cal Q}}
\def\cR{{\cal R}}
\def\cT{{\cal T}}
\def\cU{{\cal U}}
\def\cV{{\cal V}}
\def\cW{{\cal W}}
\def\cX{{\cal X}}
\def\cY{{\cal Y}}
\def\cZ{{\cal Z}}

\def\tq{{\ti q}}
\def\tta{{\ti \tau}}

\def\mC{{\mathbb C}}
\def\mZ{{\mathbb Z}}
\def\mR{{\mathbb R}}
\def\mN{{\mathbb N}}
\def\rot{{\rm curl}}
\newcommand{\li}{\lim_{n\rightarrow \infty}}
\newcommand{\mat}[4]{\left(\begin{array}{cc}{#1}&{#2}\\{#3}&{#4}
\end{array}\right)}
\newcommand{\thmat}[9]{\left(
\begin{array}{ccc}{#1}&{#2}&{#3}\\{#4}&{#5}&{#6}\\
{#7}&{#8}&{#9}
\end{array}\right)}
\newcommand{\beq}[1]{\begin{equation}\label{#1}}
\newcommand{\beqnl}{\begin{equation}}
\newcommand{\eq}{\end{equation}}
\newcommand{\beqn}[1]{\begin{eqnarray}\label{#1}}
\newcommand{\eqn}{\end{eqnarray}}
\newcommand{\p}{\partial}
\newcommand{\di}{{\rm diag}}
\newcommand{\oh}{\frac{1}{2}}
\def\1m{^{-1}}
\def\im{\Im m\,}
\newcommand{\lab}[1]{\label{#1}}

\newcommand{\su}{{\bf su_2}}
\newcommand{\uo}{{\bf u_1}}
\newcommand{\SL}{{\rm SL}(2,{\mathbb C})}
\newcommand{\slt}{{\rm sl}(2,{\mathbb C})}
\newcommand{\GLN}{{\rm GL}(N,{\mathbb C})}
\def\sln{{\rm sl}(N, {\mathbb C})}
\def\SLN{{\rm SL}(N, {\mathbb C})}
\newcommand{\gln}{{\rm gl}(N, {\mathbb C})}
\newcommand{\PSL}{{\rm PSL}_2( {\mathbb Z})}
\def\f1#1{\frac{1}{#1}}
\def\lb{\lfloor}
\def\rb{\rfloor}
\def\sn{{\rm sn}}
\def\cn{{\rm cn}}
\def\dn{{\rm dn}}
\def\wtS{\widetilde{sin}_\te}
\newcommand{\rar}{\rightarrow}
\newcommand{\upar}{\uparrow}
\newcommand{\sm}{\setminus}
\newcommand{\ms}{\mapsto}
\newcommand{\bp}{\bar{\partial}}
\newcommand{\bz}{\bar{z}}
\newcommand{\bA}{\bar{A}}
\newcommand{\bL}{\bar{L}}
\newcommand{\bG}{\bar{G}}

\def\frak{\mathfrak}
\def\gg{{\frak g}}
\def\gl{{\frak l}}
\def\ge{{\frak e}}
\def\gJ{{\frak J}}
\def\gS{{\frak S}}
\def\gA{{\frak A}}
\def\gb{{\frak b}}
\def\gd{{\frak d}}
\def\gh{{\frak h}}
\def\gn{{\frak n}}
\def\gH{{\frak H}}
\def\gM{{\frak M}}
\def\gt{{\frak t}}
\def\gx{{\frak x}}
\def\Ab{\cA d}

\newcommand{\ran}{\rangle}
\newcommand{\lan}{\langle}

\def\bw{\bar{w}}

\newcommand{\vtb}{\theta_{10}}
\newcommand{\vtc}{\theta_{00}}
\newcommand{\vtd}{\theta_{01}}

\newcommand{\sect}[1]{\setcounter{equation}{0}\section{#1}}
\renewcommand{\theequation}{\thesection.\arabic{equation}}
\newtheorem{predl}{Proposition}[section]
\newtheorem{defi}{Definition}[section]
\newtheorem{rem}{Remark}[section]
\newtheorem{cor}{Corollary}[section]
\newtheorem{lem}{Lemma}[section]
\newtheorem{theor}{Theorem}[section]

\def\teal{\te\ep\cdot\al}
\def\tega{\te\ep\cdot\ga}
\def\eal{\ep\cdot\al}
\def\ega{\ep\cdot\ga}

\newcommand{\ttbs}{\char'134}
\newcommand{\AmS}{{\protect\the\textfont2
  A\kern-.1667em\lower.5ex\hbox{M}\kern-.125emS}}

\hyphenation{author another created financial
 paper re-commend-ed Post-Script}

\vspace{0.3in}
\begin{flushright}
 IITP-TH-08/15\\
\end{flushright}
\vspace{10mm}
\begin{center}
{\Large{\bf  Geometry of Higgs bundles over elliptic curves related to automorphisms of simple Lie algebras,
 Calogero-Moser systems and KZB equations}
}\\
\vspace{12mm}
{\large {A. Levin}\,$^{\flat\,\sharp}$ \ \ \ \ {M.
Olshanetsky}\,$^{\S}$
 \ \ \ {A. Zotov}\,$^{\diamondsuit\, \sharp\, \natural}$ }\\
 \vspace{8mm}
\end{center}
\vspace{2mm}
 \S --
{\small{\sf Institute for Information Transmission Problems, 127994 Moscow, Russia}}\\
\vspace{2mm}
 $^\flat$ -- {\small{\sf 
 NRU HSE, Department of Mathematics,
 Myasnitskaya str. 20,  Moscow,  101000,  Russia}}\\
 \vspace{2mm} $^\sharp$ -- {\small{\sf 
 ITEP, B. Cheremushkinskaya str. 25,  Moscow, 117218, Russia}}\\
 \vspace{2mm} $^\natural$ -- {\small{\sf MIPT, Inststitutskii per.  9, Dolgoprudny,
 Moscow region, 141700, Russia}}\\
\vspace{2mm} $^\diamondsuit$ -- {\small{\sf Steklov Mathematical
Institute  RAS, Gubkina str. 8, Moscow, 119991,  Russia}}\\

\vspace{2mm}
\begin{center}
\footnotesize{{\rm E-mails:}{\rm\ \
 alevin@hse.ru,\  olshanet@itep.ru,\  zotov@mi.ras.ru}}
\end{center}

\begin{abstract}
We construct twisted Calogero-Moser (CM) systems with spins as the Hitchin systems derived from the
Higgs bundles over elliptic curves, where  transitions operators are defined by an arbitrary finite order
 automorphisms of the underlying Lie algebras.
 In this way we obtain  the spin generalization of the twisted D'Hoker- Phong and
 Bordner-Corrigan-Sasaki-Takasaki
 systems.
 As by product, we construct the corresponding
twisted classical dynamical r-matrices and Knizhnik-Zamolodchikov-Bernard equations related to
the automorphisms of the Lie algebras.
\end{abstract}

\tableofcontents

\section{Introduction}

\setcounter{equation}{0}

We consider a class of integrable systems similar to those that introduced some time ago in two cycles of works
\cite{BST} and \cite{DHP}.
It is so-called twisted elliptic Calogero-Moser (CM) models related to a simple group $G$.
Implicitly, they were appeared independently in the work of Etingof and Schiffmann \cite{ES}
 devoted to classification of dynamical classical $r$- matrices. This type of $r$-matrices
 were introduced in \cite {FP}.
In this paper we identify these system with a specific class of  the Hitchin system related
 to elliptic curves $\Si_\tau$ ($\tau$ is the modular parameter) and in this way associate them with special Higgs bundles over elliptic
  curves.

It was discovered in \cite{GN} (see also \cite{ER,N}) that
the standard $SL_N$ CM systems can be considered as a Hitchin system related to
the trivial Higgs bundle over an elliptic curve with a marked point.
This approach was generalized to elliptic CM systems related to arbitrary
simple Lie groups and topologically non-trivial Higgs bundles in \cite{LOSZ1,LOSZ2}. Simultaneously,
the dynamical classical $r$- matrices
were derived in the similar way via the Poisson reduction from the canonical structure
on the Higgs bundles (see also the
earlier works \cite{BDOZ,CLOZ}).
The underlying holomorphic bundles in this construction are defined by
two transition operators, satisfying some commutativity conditions. If their group commutator
is the identity element, then the bundle is topologically trivial. In this case we come to the
 elliptic CM systems with spin \cite{GH,LX,Wo}.  If the  group commutator is
equal to a non-identity element of the center$\cZ(G)$, the bundle is nontrivial.
It is described by the characteristic classes defined as an element of $H^2(\Si_\tau,\cZ(G))$.
 The moduli space of these bundles,
the corresponding elliptic CM systems and dynamical $r$- matrices were considered in
 \cite{LOSZ1,LOSZ2}.

In this paper we replace one of the transition operators
by the finite-order automorphisms of the corresponding Lie algebras.
We describe the moduli spaces of the
trivial and nontrivial holomorphic bundles obtained in this way in terms of the quotient of
Cartan algebras with respect to the action of the Bernshtein-Schwarzman group (\ref{equ}), (\ref{msa}), (\ref{mse}).
We construct the corresponding Higgs bundles and  twisted CM systems.
\footnote{For the trigonometric (Calogero-Sutherland) systems the similar approach was developed in \cite{FP1,FP2}.}
  In addition to the Lax operators and the Hamiltonians (\ref{ha}), (\ref{ha1})  we derive the classical dynamical
$r$-matrices (\ref{erm}). They govern   the Poisson
structures of the corresponding systems. In such a way we partly cover the classification
of the classical dynamical $r$-matrices \cite{ES} based on symmetries of the extended
Dynkin diagrams. The moduli spaces of the dynamical $r$-matrices are defined by the moduli
spaces of the corresponding holomorphic bundles. They were intruduced and studied in \cite{ES1}.
 The $r$-matrices related to all finite automorphisms of simple Lie algebras
coming from the trivial bundle were constructed in \cite{FP}.

In particular, for the CM systems we come to results closed to \cite{BST} and \cite{DHP}.
It should be note that the systems considered in those papers are spinless CM systems
related to a simple group $G$, while in our case we come the CM systems with spin.
They can be considered a  generalization of CM type systems described in \cite{LX}.
 To come to the spinless systems one should specify representations of the underlying Lie algebra,
 as it was done in \cite{BST,DHP}.
Another approach to the spinless systems  developed in \cite{HM} is based on the equivariant embedding of $G$ in
$GL(N)$.
Note, that the Hitchin type derivation of the twisted Calogero systems was proposed also in \cite{KT}.


Following the general prescription \cite{Fe,FW} we construct the
Knizhnik-Zamolodchikov-Bernard (KZB) connection using
the classical $r$-matrices. It is flat connection in the bundle of conformal blocks
over the moduli space of elliptic curves with marked points. The conformal blocks define
 the correlators  in the WZW theory on elliptic curves.
 The fields in the theory considered in
 this paper satisfy the specific quasi-periodicity conditions.
  The monodromy of
 the fields around one of the cycles is defined by the outer automorphism of the underlying
 simple algebra. This construction is very similar to one proposed in \cite{KuT,LOSZ3}.

 {\small {\bf Acknowledgments.}\\
 The work of M.Olshanetsky was performed at the Institute for Information Transmission Problems
 with the financial support of the
Russian Science Foundation (Grant No.14-50-00150)}


\section{General construction}

\setcounter{equation}{0}

Let $\Si_{g}$ be a Riemann surface of genus $g$, $G$ is a simple complex Lie group,
and $\gg$ its Lie algebra.
Let $\cP$ be a principal $G$-bundle over $\Si_{g}$ and $E_G=\cP\times_{G}V$
is the associated bundle ($V$ is a $G$ module).

We will pay a special attention on the case,
when $V$ is the Lie algebra $\gg$ with the adjoint action $Ad_G$ of $G$ on $\gg$.
In this case we deal with the adjoint bundle $\Ab_G=\cE_G\times_{Ad_G}\gg$.
 The fibers of the bundle carry  the Lie algebra $\gg$ structure.

 \emph{The Higgs bundle} $\cH(G,\Si_g)$ over the Riemann surface $\Si_{g}$ is the pair $(E_G,\Phi)$,
 where $\Phi$ is a Higgs field.
 It is a holomorphic 1-form on $\Si_{g}$ taking values in $End\,E_G$.

 Here we investigate the Higgs bundles over the elliptic curves (the genus one case) and
 construct the corresponding integrable systems.


 \subsection{Holomorphic bundles over elliptic curves}\label{elb}

Our first goal is to describe the moduli spaces and to classify the topological types
of these bundles, when $\Si_g$ is an elliptic curve $\Si_\tau\sim\mC/(\tau\mC\oplus\mC)$,
 $\,(Im\,\tau>0)$.

The moduli spaces $Bun(\Si_\tau,G)$ can be defined in the following way \cite{Atiyah,BS,Lo}.
Let $\rho_1$ and $\rho_\tau$ be two generators of the fundamental group
$\pi_1(\Si_\tau)$ corresponding to
the shifts $z\to z+1$ and $z\to z+\tau$ satisfying the relation
 \beq{fg}
\rho_1\rho_\tau\rho^{-1}_1\rho^{-1}_\tau=1\,.
 \eq
The sections of a $G$-bundle $E_G(V)$ over $\Si_\tau$ satisfy the
quasi-periodicity conditions
 \beq{s1}
\psi(z+1)=A(z)\, \psi(z)\,,~~~\psi(z+\tau)=B(z)\, \psi(z)\,,~~(\psi\in\G(E_G(V)))\,,
 \eq
where the transition operators $(A,B)$ should respect (\ref{fg})
 \beq{gce2}
A(z)B^{-1}(z)A^{-1}(z+\tau)B(z+1)=1\,.
 \eq
 By definition, these transition operators define a topologically trivial bundle.

Let  $\zeta$ be an element of the center  $\cZ( \bG)$ of the universal covering group $\bG$.
To come to a non-trivial bundle replace (\ref{gce2}) by equation
 \beq{gce}
A(z)B^{-1}(z)A^{-1}(z+\tau)B(z+1)=\zeta\,.
 \eq

A bundle $\ti E$ is equivalent to $E$ if its sections $\ti \psi$
are related to $\psi$ as $\ti \psi(z)=f(z)\psi(z)$, where $f(z)$ is
invertible operator acting in $V$. It follows from (\ref{s1}) that
transformed transition operators have the form:
 \beq{gtt}
A^f(z)=f(z+1)A(z) f^{-1}(z)\,,~~~~B^f(z)=f(z+\tau)B(z) f^{-1}(z)\,.
 \eq
These transformations form the group of automorphisms $\cG$ (the gauge group) of the bundle $E_G$.

According to \cite{NS} these the transition operators can be transformed to  constants
(see below Proposition\,\ref{mbu}).
Thereby, $(A,B)$ form the projective representation of $\pi_1(\Si_\tau)$.
 The moduli space of stable holomorphic bundles over  $\Si_\tau$ can be defined as
     \beq{mhol}
    \fbox{$
 Bun_\zeta(\Si_\tau,G) =\{[A,B^{-1}]=\zeta\}/\cG\,,~~\zeta\in\cZ(G)\,.$}
     \eq
It was proved in \cite{BS,Lo} that $Bun_\zeta(\Si_\tau,G)$ for $\zeta=1$ is isomorphic to a weighted
projective space.

 Consider the universal covering group $\bG$ and the quotient group $G^{ad}=\bG/\cZ(\bG)$,
where $\cZ(\bG)$ is the center of $\bG$
\beq{cen}
1\to\cZ(\bG)\to\bG\to G^{ad}\to 1\,.
\eq
Then for the analytic sheaves over $\Si_\tau$  we write the exact sequences
$$
  1\to\cZ(\bG))\to\bG(\cO_\Si)\to G^{ad}(\cO_\Si)\to 1\,, \\
$$
Then we come to the long exact sequences
\beq{coh1}
 H^1(\Si_\tau,\cZ(\bG))\to H^1(\Si_\tau,\bG(\cO_\Si))\to H^1(\Si_\tau,G^{ad}(\cO_\Si))
 \to H^2(\Si_\tau,\cZ(\bG))\sim\cZ(\bG))\to 0 \,,
\eq
The first cohomology group $H^1(\Si_\tau,G(\cO_\Si))$  of $\Si_\tau$ with coefficients in analytic sheaves
defines  the moduli space $ Bun_\zeta(\Si_\tau,G)$ of the $E_G$-bundles.
The elements from $H^2(\Si_\tau,G)$ are obstructions to lift the $G^{ad}$ bundles to $\bG$ bundles.
They provide the topological classification of the holomorphic vector bundles
$$
\zeta(E_{G^{ad}})\in H^2(\Si_\tau,\cZ(\bG)) -{\rm ~obstructions ~to~ lift~}
 E_{G^{ad}}-{\rm bundle~to~}E_{\bG}-{\rm bundle}\,.
$$
\begin{defi}
Images of $H^1(\Si_\tau,G(\cO_\Si))$ in $H^2(\Si_\tau,\cZ)$ are called the characteristic classes
$\zeta(E_G)$ of $G$-bundles.
\end{defi}

The cotangent bundle  $T^*Bun(\Si_g,G)$ to the moduli space is the
moduli space of the Higgs bundle
over $\Si_g$. It is obtained from $\cH(G,\Si_g)$ via the symplectic reduction.
  It is the phase space of algebraically integrable system \cite{Hi}. In the
case $T^*Bun_\zeta(\Si_\tau,G)$ we obtain the Calogero-Moser type systems \cite{GN,LOSZ1,LOSZ2}.
To derive in this way the twisted Calogero-Moser systems and twisted
Knizhnik-Zamolodchikov-Bernard equations we replace the transition operator $A$ (\ref{s1}) on
an outer automorphism of the underlying Lie algebra.

%


\section{The finite order automorphisms of simple Lie algebras}

\setcounter{equation}{0}

\subsection{Inner and outer automorphisms}

The group of finite order automorphisms of simple complex Lie algebra $\gg$ is a semidirect product of
the subgroup of
finite order inner automorphisms $Inn(\gg)$  and the outer automorphisms $Out(\gg)$
$$
1\to Inn(\gg)\to Aut(\gg)\to Out(\gg)\to 1\,.
$$
Therefore, the outer automorphisms form the quotient group
$Out(\gg)=Aut(\gg)/Inn(\gg)$.

Let $G$ be a simple complex Lie group and Lie$(G)=\gg$.
The outer automorphism group $Out(G)$ of the group $G$ is the quotient $Aut(G) / Inn(G)$,
where
$$
Inn(G)=\{\iota_h\,g\to hgh^{-1}\}\sim G^{ad}
$$
 is the subgroup consisting of inner automorphisms.
The outer automorphism group $Out(G)$ is dual to the center $\cZ(G)$ in the following sense.
The inner automorphism $\iota $ of $G$ leads to  the short exact sequence:
$$
\cZ(G)\hookrightarrow G\stackrel{\iota}\rightarrow Aut_G\rightarrow Out_G\,.
$$
Here $\cZ(G)=Ker(\iota)$ and $Out(G)\sim Cokernel(\iota)$.

The $Aut(\gg)$ action has the following description \cite{Ka,Ka1,VO}.
Let us fix a Cartan subalgebra $\gh\subset\gg$ and system of simple roots $\Pi=\Pi(\gg)$.
Define group  $Aut(\gg,\gh,\Pi)$ of automorphisms preserving $\gh$ and $\Pi$.
Let $H^{ad}=\exp ad\,(\gh)$ be a maximal torus in the group $G^{ad}$ and $Aut(\Pi)$ is a group of
symmetries of the Dynkin diagram corresponding to $\Pi$.
Then
\beq{sdd}
Aut(\gg,\gh,\Pi)=H^{ad}\ltimes Aut(\Pi)\,,
\eq
and the general automorphisms are represented in the form
$Aut(\gg)=Aut(\gg,\gh,\Pi)\cdot Inn(\gg)$.

Our goal is to construct a basis in $\gg$ compatible with the $Aut(\gg)$ action.
It is presented in Appendix C.


\subsection{Generic finite order automorphisms automorphisms }

In this subsection we follow \cite{Ka,Ka1} (see also Appendix C).

We define the general automorphism of order $p$ using (\ref{sdd}).
Let $\gg_0$ be the invariant subalgebra of $\nu$ (\ref{rdi}) and $\gh_0$ its
Cartan subalgebra.
Using this structure we can represent the automorphism
 $\si\in Aut(\gg,\gh,\Pi)$ (\ref{sdd}) in the form
\beq{rga}
\si=\nu\exp\left(\frac{2\pi\imath}p ad_\ka\, \right)\,,~~\ka\in\gh_0\,.
\eq

Let $(s_0,s_1,\ldots,s_n)$ be nonnegative integers without nontrivial common factor and
\beq{oa}
p=r\sum_{j=0}^na_js_j\,,
~~(\nu^r=Id)\,,
 \eq
 (see (\ref{dea})). The general automorphism $\si_{r,s}$ of $\gg$ is defined by its action on the root subspaces
$(E_{\ti\al_k}\,,~\ti\al_k\in\ti\Pi\,,~(\ref{isr})\,, ~E_{-\ti\al_0}\,,~(\ref{dea}))$
\beq{ina}
\si_{r,s}E_{\ti\al_k}=\bfe(rs_k/p)E_{\ti\al_k}\,,~~k=0,\ldots,n\,.
\eq
If $\nu=Id$ it is the adjoint action of the invariant torus
$H_0^{ad}=\exp\left(\frac{2\pi\imath}p ad_\ka\right)$\,, $\,(\ka\in\gh_0)$ on $\gg$.
Define the element $\kappa\in\gh_0$ by its expansion in the coroots basis $H_j$ (\ref{pid})
$$
\kappa=\sum_{j=1}^nc_jH_j\in\gh_0 \,,
$$
 such that Ad$_{\bfe(\kappa)}=\si_{1,s}$. Then the coefficients $c_j$ are defined as follows.
Let $(\ti a)_{jk}=\lan\ti\al_k,H_j\ran$ be the
Cartan matrix corresponding to the twisted root system $\ti\Pi$ (\ref{isr}).
Then  $c_j=(\ti a^{-1})_{jk}s_kr/m$ and
\beq{cq1}
\cQ=\bfe(\kappa)=\exp\left(2\pi\imath\frac{r}p\sum_{j,k=1}^n(\ti a^{-1})_{jk}s_kH_j\right)\,.
\eq
Then
$Ad_{\cQ}E_{\ti \al_k}=\bfe\left(\frac{r}ps_k\right) E_{\ti \al_k}$.
It follows from (\ref{dea}) and (\ref{oa1} ) that $\bfe(\frac{r}p\sum_{k=1}^na_ks_k)=
\bfe(-\frac{r}pa_0s_0)$. Thus,
\beq{akm}
\si_{1,s}E_{-\ti\al_0}\equiv Ad_{\cQ}E_{-\ti \al_0}=\bfe(s_0r/p)E_{-\ti\al_0}
\eq
and in this way $\si_{r,s}$ really defines the automorphism of  order $p$.
It is generated by the action (\ref{rga})
\beq{efs}
\si_{r,s}=\nu\cQ=\nu\bfe(\kappa)\,,~~\nu\in Out(\gg)\,,  ~~\nu^r=Id\,.
\eq
The automorphism $\si_{r,s}$ defines $\mZ/p\mZ$-gradation of the Lie algebra $\gg$
\beq{gra}
\gg=\sum_j\gg_j\,,~~\si_{r,s}(\gg_j)=\om\gg_j\,,~~\om^p=1\,.
\eq


\subsection{Loop algebras and loop group}

Let $\cL(\gg,t)=\gg\otimes C[t^{-1},t]]$ be the loop algebra. It is a map $\mC^*\to\gg$.
Define  the action of the  automorphism $\si_{r,s}$ on $t$ as $\si_{r,s}(t)=\om t$.
 Consider equivariant maps $\cL_0(\gg,t,\si_{r,s})\in\mC^*\to\gg$ with respect to $\si_{r,s}$
\beq{ins}
\cL_0(\gg,t,\si_{r,s})=\cL_0(\si_{r,s}(\gg),\om^{-1}t)\,,~~
\cL_0(\gg,t,\nu)=\sum_jt^j\gg_{j,\,mod\, p}\,,
\eq
where $\gg_{j}$ is defined in (\ref{gra}).


Let $L(G)$ be the loop group corresponding to the $G$
  \beq{logr}
L(G)=G\otimes C[t^{-1},t]]=\left\{\,\sum_kg_kt^k\,,~g_k\in G\,\right\}\,.
  \eq
Let $N^-$ be the negative unipotent subgroup of $G$ and $B$ is the positive Borel subgroup.
Define  the loop subgroups
    \beq{bpl}
L^+(G)=\{g_0+g_1t+\ldots=g_0+tL[t]]\}\,,~~g_j\in G\,,~g_0\in B\,,
   \eq
    \beq{bne}
N^-(G)=\{n_-+g_1t^{-1}+\ldots=n_-+t^{-1}L[t^{-1}]\}\,,~~n_-\in N^-\,.
   \eq
   Consider two types of the affine Weyl groups:
  \beq{awg}
W_P=\{\hat w=wt^\ga\,,~w\in W\,,~\ga\in P^\vee\}\,,~~W_Q=\{\hat w=wt^\ga\,,~w\in W\,,
~\ga\in Q^\vee\}\,.
   \eq
 They act on the root vectors as
 $ E_{\hat\al}= E_\al t^n\to E_{\hat w(\hat\al)}=E_{w(\al)}t^{n+\lan\ga,\al\ran}$.
For the loop groups we have the affine Bruhat decompositions \cite{PS}
\beq{PS}
 L(G^{ad}) =\bigcup_{ w\in\ti  W_P}N^-(G^{ad})\hat wL^+(G^{ad})\,,
\eq
\beq{PS1}
   L(\bG) =\bigcup_{\hat w\in W_Q}N^-(\bG)\hat wL^+(\bG)\,.
      \eq


Consider  the twisted loop group \cite{PR} with the Lie algebra
$\cL_0(\gg,t,\si_{r,s})$ (\ref{ins}). It is the subgroup of  the equivariant maps
\beq{le}
L_0(G,\si_{r,s})=\{g(t)\,|\,\si_{r,s}(g(t)=g(\om t)\}\,.
\eq
Let us look on its decompositions (\ref{PS}), (\ref{PS1})
\beq{geq}
L_0(G,\si_{r,s})=\{g(t)=g_-(t)t^\ga g_+(t)\}\,,~~
\left\{
\begin{array}{cl}
1.~  \ga\in \ti Q^\vee\,,~(\ref{cwl2}) & \mbox{for} ~G=\bG\,, \\
 2.~ \ga\in \ti P^\vee\,, ~(\ref{cwl1})& \mbox{for} ~G=G^{ad}\,.
\end{array}
\right.
\eq
 From (\ref{le}) we find that $\si_{r,s} (g_\pm(t))=g_\pm(\om t)$ and
 \beq{aul}
 \om^\ga t^\ga=t^{\si_{r,s}(\ga)}\,.
 \eq
 From this condition we find that $\ga$  are not arbitrary,
  but satisfies the following conditions
 \beq{cl}
(\om^\ga=1\,,~t^\ga=t^{\si_{r,s}(\ga)})\leftrightarrow
\left\{
\begin{array}{cl}
1.~  \ga\in p\ti Q^\vee & \mbox{for} ~G=\bG\,, \\
 2.~ \ga\in p\ti P^\vee & \mbox{for} ~G=G^{ad}\,,
\end{array}
\right.
~(p=ord\,\si_{r,s})\,,
 \eq
 where $\ti P^\vee$ and $\ti Q^\vee$ are invariant coweight and coroot lattices (\ref{cwl1}),
 (\ref{cwl2}).



\section{Holomorphic bundles over elliptic curves and automorphisms of Lie algebras}

\setcounter{equation}{0}

 We construct holomorphic bundles over elliptic curves as in Section\,\ref{elb} passing to
the Jacobi description of the elliptic curves as the quotient of $\mC^*$.
Let $t=\exp\,2\pi\imath z/p=\bfe(z/p)\in\mC^*$, where $p=$ord$\,\si$, be the parameter of the
loop groups and algebras considered in previous Section.
The action on $z$ of the lattice $\mZ\oplus\tau\mZ$ in terms of $t$ assumes the form
$$
\begin{array}{cl}
  z\to z+1 & t\to t\om\,,~(\om=\bfe(1/p))\,, \\
  z\to z+\tau & t\tq\,,~(\ti q=\bfe(\tta)\,,~\,,\tta=\tau/p)\,.
\end{array}
$$
In the coordinate $t$ the
curve $\Si_\tau$ is defined as the quotient
$ C_{\ti q}=\mC^*/\mZ^{\ti q}$.
We identify $C_\tq$ with the annulus
\beq{cq}
C_\tq=\{t\in\mC^*\,|\,|\tq^\oh|\leq |t|< |\tq^{-\oh}\}\,.
\eq



\subsection{Adjoint bundles}

Let $\cP$ be a principle $G$-bundle over $C_\tq$.
Consider the bundle $\Ab_G^\si=\cP\otimes_{G^{ad}}\gg$  over $C_q$. Its
 sections   $\G(\Ab_G^\si)=\{\Psi(t)\}$ are the equivariant holomorphic maps $ C^*\to\gg$
with respect to the action of
finite order automorphism $\si_{r,s}$ (\ref{ins}).
\beq{tw1}
\Psi(t\om)=\si_{r,s}(\Psi(t))\,,~~\om^p=1\,.
\eq
The transition operator corresponding to another cycle is
\beq{tw2}
\Psi(\tq t)=R(t)\Psi(t)\,,
\eq
where the $R(t)$ action is the adjoint action, $R(t)\in L(G^{ad})$ (\ref{PS}) for $\hat w=1$.

The consistency condition of these transformations implies
\beq{cc1}
\si_{r,s}( R(t))=\zeta R(t\om)\si_{r,s}\,,~~\zeta\in\cZ(\bG)\,.
\eq
If $\zeta=Id$,  then
(\ref{cc1}) means that $R(t)\in L_0(G^{ad},\si)$ (\ref{le}).

The gauge group $\cG_{\si}(G)$ is the group of automorphisms  of the $\Ab_G^\si(C_\tq)$-bundle.
It follows from  (\ref{tw1}) that it is the group of equivariant  holomorphic maps (\ref{le}).
Remind that in the decomposition (\ref{geq}) $\ga$ is not arbitrary but satisfies
(\ref{cl}.2).

The group $\cG_{\si}(G)$ acts on the  transition operators as
 \beq{gtj}
R(t)\to f(\tq t)R(t)f^{-1}(t)\,.
\eq
This action preserves (\ref{cc1}).

 We attach  to the marked point $t=1$ a
flag variety $Flag\sim G/P$, where $P$ is a parabolic subgroup of $G$ and assume
that $\cG_{\si}(G)$ preserves the flag. Then the gauge group is defined as
\beq{gga}
\cG_{\si}(G)=\{f(t)\,:C^*\to G^{ad}\,, ~f\in L_0(G^{ad},\si)\,,~f(t)|_{t=1}=P)\}\,.
\eq

\begin{lem}\lab{l1}
The bundle $\Ab_G^\si$ defined by (\ref{tw1}), (\ref{tw2}) is gauge equivalent to the bundle
$\Ab_G^\nu$, where $\nu$ is the diagram automorphism.
$\Ab_G^\nu$ is defined by the transitions of its sections
\beq{tt1}
\ti\Psi(t\om)=\nu(\ti\Psi(t))\,,~~\om=\bfe(1/r)\,,
\eq
\beq{tt2}
\ti\Psi(\tq t)=\ti R(t)\ti\Psi(t)\,,~~ \ti R(t)= Ad_{t^{-p\kappa}}R(t)\,,
\eq
where $p=ord\,\si$ and $\bfe(\ka)$ is (\ref{cq1})).
\end{lem}

\emph{Proof }\\
Consider the gauge transformation
$f(t)=t^x$, where $x\in\gh_0$. It follows from (\ref{tw1})  that
$$
\si_{r,s}\stackrel{f}{\longrightarrow} \om^xt^x\si_{r,s}t^{-x}=\om^x\si_{r,s}\,.
$$
If $x=-p\kappa$, where
$\kappa$ is defined in (\ref{cq1}), then under this transformations
$\si_{r,s}\stackrel{f}{\longrightarrow}  \nu$ and
 we replace
 (\ref{tw1}) on (\ref{tt1}).

In this way we stay with the diagram automorphism (\ref{tt1}). Acting by
\beq{tka}
t^{-p\kappa}=t^{r\sum_{j,k=1}^n(a^{-1})_{jk}s_kH_j}\,,~~
\left(\kappa=\frac{ r}p\sum_{j,k=1}^n(a^{-1})_{jk}s_kH_j\right)
\eq
 on $R(t)$ we come to (\ref{tt2}).$\Box$

 \begin{rem}
Consider the  monodromy
 $\cM=\bfe(-r\sum_{j,k=1}^n(a^{-1})_{jk}s_kH_j)$ of the
gauge transform $t^{-p\kappa}$.
 The inverse Cartan matrices
$(a_{jk})^{-1}$ for  Dynkin diagrams  with nontrivial symmetries have
fractional matrix elements $(a_{jk})^{-1}$ such that $(a^l_{jk})^{-1}$ are integer,
 where $l$ is the order of the center $\cZ(\bG)$ (\ref{cen}) of $\bG$. Since the center
 is isomorphic to the quotient of the  weight and the root lattices $\cZ(\bG)\sim P/Q$,
 the monodromy $\cM$ is the element of $\cZ(\bG)$. Therefore, $\cM$ is well defined
 as an element of the Cartan subgroup $ H^{ad}\subset G^{ad}$.
 \end{rem}


 \begin{rem}
It follows from (\ref{tt1}) that the sections $\ti\Psi\in\G(\Ab_G^\nu)$  have the periodicities
\beq{np}
\ti\Psi(t\om^{r})=\ti\Psi(t)\,,~~(\om=\bfe(1/p))\,,
\eq
while the sections $\Psi\in\G(\Ab_G^\si)$  have the periodicities
$$
\Psi(t\om^{ p})=\Psi(t)\,,~~(p\geq r)\,.
$$
\end{rem}
%


\bigskip

Now we consider the bundles $\Ab_G^\nu(C_q)$ defined by (\ref{tt1}), (\ref{tt2}).
The condition (\ref{cc1}) assume the form
\beq{cc2}
\nu( R(t))=\zeta R(t\om)\,,~~\zeta\in\cZ(\bG)\,, ~~\om=\bfe(1/r)\,.
\eq
The moduli space of the bundles for fixed
$\nu$ and $\zeta$ is defined as the set of $\{R(t)\}$ satisfying (\ref{cc2}) up the
gauge transformations (\ref{gtj})
\beq{bg}
Bun_{\nu,\zeta}(\Ab_G^\nu(C_\tq))=\{ R(t)\,|\,\nu( R(t))=
\zeta R(te^{2\pi\imath/r})\}/\cG_{\nu}(G)\,.
\eq
Here $\cG_{\nu}(G)$ is the group (\ref{gga}).
Our goal is to describe a big cell in this space.
First we prove that the bundles $\Ab_G^\nu(C_q)$ are topologically trivial.

We will prove below (Proposition\,\ref{mbu}) that any transition operator $R(t)$
 in a neighborhood of constant maps from $C_\tq$ to the Cartan subgroup
 $\bfx=\bfe(\bfu)\in\cH(G)$ is represented by the gauge transformation $f(t)=1+\gx(t)$
 \beq{fcws}
R(t)=f^{-1}(qt)\bfx f(t)\,.
\eq

 Passing to the Cartan subalgebra we rewrite (\ref{cc2}) as
\beq{cco1}
\nu(\bfu)-\bfu=\xi\,,~~(\zeta=\bfe(\xi))\,,
\eq
Since $R(t)$ takes values in $G^{ad}$ the element $\bfu$ is defined up to the shift on
\emph{ the weight lattice}
\beq{sp}
\bfu\sim\bfu+\ga\,,~~\ga\in P\,.
\eq
On the other hand, $\xi\in P$ in (\ref{cco1}) is defined up to addition an element from
\emph{the root lattice }$\be\in Q$. Note that $\xi$ is not an arbitrary element of $P$
 but belongs to the image of the operator $\Upsilon$ acting on $\gh$ as
\beq{ip}
\Upsilon\,:\gh\to\gh\,,~x\mapsto \nu(x)-x\,.
\eq
In particular, it acts on the weight lattice $\Upsilon\,:\,P\to P$.
Let
\beq{pip}
P^\Upsilon= im\,(\Upsilon)\,,~P^\Upsilon=\{\xi\in P\,|\,\xi=\nu(\ga)-\ga\,,~\ga\in P\}\,.
\eq

\begin{predl}
The bundles $\Ab^\nu_G(C_q)$ are topologically trivial.
\end{predl}
\emph{Proof}\\
We prove that any $\xi\in P^\Upsilon$ (\ref{pip}) can be gauged away
 by the shift of $\bfu$ (\ref{sp}) by an element $\ga\in P$
 $$
 \left(\xi\in\, im\,(\Upsilon)\right)\Rightarrow\left(\xi=\nu(\ga)-\ga\,,~\ga\in P\right)\,.
 $$
  Represent $x\in im\,(\Upsilon)$
 in the basis $\Xi$ of fundamental weights (\ref{fw}) $x=c_1\varpi_1+\ldots+c_n\varpi_n$. Since
 $\nu(\Xi)=\Xi$  then $\Xi$ is an union of the orbits
$$
 \cO_j=\{\varpi_j,\nu(\varpi_j),\ldots,\nu^{r-1}(\varpi_j)\}\,.
 $$
 of length $r=1,2$ or $3$ (for $D_4$).
 The operator $\Upsilon$ acts on  orbit $\cO$ of the length $r>1$ as the $r$-order matrix
 $$
   \left(
     \begin{array}{rrrrr}
       -1 & 1 &0&\ldots& 0 \\
       0 & -1 & 1&\ldots& 0 \\
       \vdots  &\ddots  & \ddots&\ddots&\vdots\\
       0& & &-1&1\\
       1 & 0&\ldots& 0& -1 \\
     \end{array}
   \right)\,,
$$
This operator can be inverted on the $r-1$-dimensional subspaces
 \beq{imi}
 \{c_j\varpi_j+\ldots +c_{\nu^{r-1}j}\nu^{r-1}(\varpi_j)\,|\,c_{j}+\ldots+c_{\nu^{r-1}j}=0\}\,.
\eq
  In terms of the $\nu$-action this condition can be reformulated
 as
 \beq{tbi}
\left\{ x\in im\,(\Upsilon)\right\}\Leftrightarrow \left\{x+\nu(x)+\ldots+\nu^{r-1}(x)=0\right\}\,.
 \eq

  For $\xi\in P$ all coefficients
$c_j=n_j$ are integer, and $\xi\in im\,(\Upsilon)$ if $n_{j}+\ldots+n_{\nu^{r-1}j}=0$.
In this case $\xi$ in (\ref{cco1}) can be adsorbed by the transformation $\bfu\to\bfu+\ga$, where
$\ga=\sum_jn_{j}\varpi_{j}$ and $\sum_jn_{j}=0$.$\blacksquare$\\

The last statement means that the adjoint bundles of Lie algebras defined by (\ref{tw1})
has only one component corresponding to the trivial characteristic class and in this way
the bundles are topologically trivial. Thus, instead (\ref{cco1}) we can consider
\beq{cco}
\nu(\bfu)-\bfu=0\,.
\eq
It means that $\bfu\in\gh_0$.

Now we prove that  in a neighborhood of $\bfu$ almost all
transitions $R(t)$ are images of $\bfe(\bfu)$ under small gauge transforms.
\begin{predl}\lab{mbu}
Let $\bfx=\bfe(\bfu)$,  $\bfu\in \gh_0$ (\ref{rdi}). Then there exists a
small gauge transformation ($f(t)=1+\gx(t)$\,, $\gx(t)\in Lie(\cG_{\nu}(G))$
such that near the Cartan element $\bfx$  almost all $R(t)$  can be represented as a gauge transform of $\bfx$
\beq{gfi}
R(t)=(1-\gx(tq))\bfx (1+\gx(t)) \,.
\eq
\end{predl}
\emph{Proof}\\
Represent $R(t)$ as $R(t)=\bfe(\bfu)+\de R(t)$, where $\de R(t)$ is small.
 Then infinitesimally (\ref{gfi}) can be rewritten as
\beq{opg}
\gx(t\tq)\bfe(\bfu)-\bfe(\bfu)\gx(t)=\de R(t)-\de R_0\,.
\eq
Here $\de R_0$ is a constant element. The substraction of $\de R_0$ provides the r.h.s. of
(\ref{opg}) belongs to  the image of the operator
$$
D_\tq(\bfu)\,:Lie\,(\cG_{\nu}(G))\to Lie\,(\cG_{\nu}(G))\otimes\Om^{(1,0)}(C_\tq)\,,
$$
$$
(  D_\tq(\bfu)\gx)(t)= \left(Ad_{\bfe(\bfu)}\gx(t\tq)-\gx(t)\right)dt/t\,.
$$
The prove is based on the existence of the Green function $G(\bfu,t)$ of $D_\tq$ taking values in
$\gg\otimes\gg$
\beq{grf}
G(\bfu,t\tq,t_1))\bfe(\bfu)-\bfe(\bfu) G(\bfu,t,t_1)=\cI\cdot\de(t,t_1)-\cI^0\,.
\eq
Here
conjugations by $\bfe(\bfu)$ is taken with respect to the first argument,
$\de(t,t_1)$ is a linear functional on the space of Laurent polynomials $\mC[t^{-1},t]$
(\ref{de1}), (\ref{def}) and $\cI$ and $\cI^0$ are the two-tensor on $\gg\otimes\gg$,
defined in the following way. Let $\{T_a\}$ be the basis (\ref{tb}) and $c_{a,b}=(T_a,T_b)$
is the Killing metric (\ref{blf}), (\ref{kfo}) in this basis. Then
$$
\cI=\sum_{a,b}c^{a,b}T_a\otimes T_b\,,~~\cI^0=\sum_{k=1}^l\ge^0_k\otimes \ge^0_k
\,,~(\mbox{see}\,\,(\ref{inv5})\,,\,(\ref{ocb}))\,.
$$
We have  $(\cI,T_c)_2=T_c\otimes Id$, where $(\cI,T_c)_2=\tr_2(\cI,T_c)$.
In fact,
$G(\bfu,t,t_1)$ is defined in terms of the classical $r$-matrix calculated below
(\ref{rm1}), (\ref{rm2})
$$
G(\bfu,t,t_1)=r(\bfu,t/t_1)\,.
$$
It satisfies (\ref{grf}) due to (\ref{rm1}), (\ref{rm2}), (\ref{qp3}).

Solutions of (\ref{opg}) in terms of $G(t,s)$ assume the form
$$
\gx(\bfu,t)=\oint_{|t_1|=|t|(1+\ve)}\left(G(\bfu,t,t_1)X(t_1)\right)_2
\,,~~(\ve>0)\,.
$$
where $X(s)=\de R(s)-\de R_0$.  $\blacksquare$



\bigskip

Thus,  we found a gauge transformation
\beq{gtr}
R(t)\stackrel{f(t)} \rightarrow \bfx\in \cH_0^{ad}\subset G_0^{ad}\,,~~\bfx=\bfe(\bfu)\,.
\eq
where the element $\bfu\in\gh_0$.

We analyze solutions of (\ref{cco}) up to the transformations preserving its form
(the residual gauge transformations).
Such solutions define the moduli space $Bun_{\nu,\zeta}(\Ab_G^\nu(C_q))$ (\ref{bg}) for
$\zeta=Id$.

 First we described the residual gauge transformations.
Let $W_0$ be the Weyl subgroup of $\gg_0$ defined in Appendix C.
It is the Weyl group generated by the reflections of the root system $\ti R$, and let
$\ti P^\vee$ is the coweight lattice of $\gg_0$ (\ref{cwl1}).

\begin{predl}\label{ds}
The equation (\ref{cco}) is invariant with respect the following transformations
\beq{gtr1}
  \bfu\to w\cdot(\bfu)+\ga_1+\ga_2\tau\,, ~~ \ga_1\in\ti P^\vee\,,~
  \ga_2\in \ti P^\vee\,,~w\in W_0\,, \\
\,.
\eq
\end{predl}
 \emph{Proof}\\
Remind that the elements $\exp\,2\pi\imath\ga$ for $\ga\in \ti P^\vee$ generates the center of the
universal covering group $\bG_0$. In other words, $\exp\,2\pi\imath\ga$ is  the unity element
in $\cH_0^{ad}$.
Therefore, $\Re e\,\bfu$ in (\ref{gtr}) is an element of the quotient
\beq{gad}
\Re e\,\bfu\in \gh_0^\mR/\ti P^\vee\,,~~(\bfu+\ga\sim\bfu\,,~{\rm for~}\ga\in \ti P^\vee)\,.
\eq
In this way
$\Re e\,\bfu$ can be chosen from the fundamental domain in $\gh_0^\mR$ under the action of
the affine Weyl group
$\ti W'_{aff}=W_0\ltimes\ti P^\vee$ (\ref{wap}).


The residual gauge transformations of $\bfu$ is the subgroup
 $L_0(\cH^{ad})\subset \cG_{\nu}(C_q, G^{ad})$ (\ref{gga}) of the equivariant maps
 to $\cH^{ad}$ (see  (\ref{cl}))
 \beq{rea}
 L_0(\cH^{ad})=\{f(t)=t^\ga h(t)\,|\,\ga\in r\ti P^\vee\,,~h(t)=\sum_kh_kt^k\,,~h_k\in\cH^{ad}\,,
 ~\nu(h_k)=\om^{-k}h_k\}\,.
 \eq
It  acts on
$\bfx$ (\ref{gtr}) and on $\bfu$ as
\beq{acu}
\bfx\to \tq^{r\ga}\bfx=q^\ga\bfx\,,~~(\tq=\bfe(\tau/r)\,,~  q=\bfe(\tau))\,,~~
\bfu\to \bfu+\ga\tau\,,~~\ga \in \ti P^\vee\,.
\eq



Then due to (\ref{gad}) and (\ref{acu}) solutions of (\ref{cco}) are invariant under the shifts
\beq{si}
\bfu\to \bfu+\ga_1+\ga_2\tau\,,~(\ga_1\,,\,\ga_2\in \ti P^\vee)\,.
\eq
In this way, $\bfu$   belongs to the fundamental domain
\beq{bs1}
\bfu\in\gh_0/\ti W'_{BS}\,,
\eq
 where $\ti W'_{BS}$ is the Bernstein-Schwarzman group \cite{BS}generated by $W_0$ and the
 shifts (\ref{si})
 \beq{BS}
\ti W'_{BS}=W_0\ltimes( \tau \ti P^\vee \oplus \ti P^\vee)\,.
\eq
 $\blacksquare$

Thus the moduli space $Bun_{\nu,Id}(\Ab_G^\nu(C_\tq))$ (\ref{bg}):
\begin{itemize}
\item is independent on $\zeta$ (\ref{bg});
\item  is isomorphic to the fundamental domain of the Bernstein-Schwarzman group (\ref{BS})
\beq{equ}
\fbox{$Bun_{\nu,Id}(\Ab_G^\nu(C_\tq))\sim\gh_0/\ti W'_{BS}(r)\,.$}
\eq
\end{itemize}
It can be proved that this quotient can be described in terms of the  curve $C_{q}$ and
the invariant weight sublattice $\ti P^\vee$
\beq{equ1}
\gh_0/\ti W'_{BS}(r)\sim (C_{q}\otimes\ti P^\vee)/W_0\,,~~q=\bfe(\tau )\,.
\eq

To pass from the $\Ab^\nu_G(C_\tq)$-bundles to $\Ab^\si_G(C_\tq)$-bundles
 we  act on the sections $\G(\Ab^\nu_G(C_\tq))$ by the gauge transform $t^{p\kappa}$ (\ref{tka}).


%



\subsection{Holomorphic bundle with non-diagonalized automorphisms}

Here we consider a holomorphic bundle (\ref{tw1}), (\ref{tw2}), where $R(t)$ is not
 an element of the Cartan subgroup $\cH(G)$
as was assumed in (\ref{fcws}) and Proposition\,\ref{mbu}, but belongs to the Weyl group $W(G)$.
More concretely, let $G=\SLN$, $N=2n$ and $\nu$ is the outer automorphism of the Lie algebra $\sln$ (\ref{alge})
leading to the invariant subalgebra sp$(n)$.
 The action of $R(t)$
is generated by the automorphism $\la$ of the extended Dynkin diagram (\ref{edg}) $\ti\Pi^{ext}$ of
the root system $A^{(1)}_{2n-1}$. It acts
 by the cyclic permutation on the roots $\al_0\to\al_1\to\al_2\to\cdots\to\al_0$, and $R(t)=\La=\la^n$.
  Represent  $x\in\sln$ as the $2n\times 2n$ matrix
\beq{sln}
x=
\left(
  \begin{array}{cc}
    A & B \\
    D & C \\
  \end{array}
\right)\in\sln\,,
\eq
Define the action of the outer automorphism $\nu$ as
\beq{an}
\nu(x)=
\left(
  \begin{array}{cc}
    -\ti D & \ti B \\
    \ti C & -\ti A \\
  \end{array}
\right)\,,~~\ti A=jA^Tj\,,~~j=\{E_{jk}=\de_{j,n-k+1}\}\,.
\eq
The action of $\la$ is performed by the conjugation by $\La$
\beq{lam}
\La=
\left(
  \begin{array}{cc}
    0 & Id_n \\
    Id_n & 0 \\
  \end{array}
\right)\,.
\eq
Note that the transformations $\nu$ and $\la$ anti-commute
\beq{cc3}
\nu(\La)\La^{-1}=-Id_N\,.
\eq

\subsubsection{Special basis in $\sln$}

Since $\nu^2=Id$ and $\La^2=Id$ and due to (\ref{cc3}) the Lie algebra $\sln$ (and more generally
$\gln$) acquires
  $\mZ_2\oplus\mZ_2=\{(a,b)\}$ gradation
$$
Ad_{\La}x=\pm x\,,~~\nu(x)=\pm x\,,
$$
\beq{ta1}
\begin{tabular}{|c|c|}
  \hline
  $(a,b)$ & $(\La,\nu)$\\
  \hline
  (0,0) & $(+,+)$ \\
  (0,1) & $(+,-)$ \\
  (1,0) & $(-,+)$ \\
  (1,1) & $(-,-)$ \\
  \hline
\end{tabular}
\eq

Instead of the $\nu$-consistent basis (\ref{bal}), (\ref{inv1}) in $\sln$ we consider
a basis related to the gradation (\ref{ta1}) and the structure (\ref{sln}).
 Consider the orbits in the space of the $\gln$-basis $\{E_{JK}\}$ under
 the action of the $\mZ_2\oplus\mZ_2$ group generated by $\nu$ and $Ad_{\La}$.
 Let $\cO_{jk}$ be the orbit passing through $E_{jk}$, or $E_{j,n+k}$ $(j,k=1,\ldots,n)$. Then
 $$
 \{E_{J,K}\,,~(J,K=1,\ldots,N)\}=\cup_{j+k\leq n+1}\cO_{jk}\,.
 $$
 $$
\sharp\,(\cO_{jk})=\left\{
\begin{array}{cc}
  4 \,,& j+k< n+1\,, \\
  2\,, & j+k=n+1\,,
\end{array}
\right.
$$

We introduce in $\gln$ the basis
\beq{gba}
\{\cT^\al_{(a,b),jk}\,,~\al,a,b=0,1\,,~j,k=1,\ldots,n\}
\eq
in terms of the tensor product of the sigma-matrices $\si_k$, $(k =0,\ldots,3)$
with the gl$(n)$ matrices. Let $E_{jk}$ be the basis in gl$(n)$.
then $\cT^\al_{(a,b)}$ is defined as follows.
For $j+k< n+1$ it takes the form
\beq{nba}
\begin{array}{cc}
\cT^0_{(0,a),jk}=\si_0\otimes(E_{jk}-(-1)^aE_{n-k+1,n-j+1})\,,&\cT^1_{(0,a),jk}=
\si_1\otimes(E_{jk}+(-1)^aE_{n-k+1,n-j+1})\,, \\
\cT^0_{(1,a),jk}=\si_3\otimes(E_{jk}-(-1)^aE_{n-k+1,n-j+1})\,,&\cT^1_{(1,a),jk}=
\si_2\otimes(E_{jk}+(-1)^aE_{n-k+1,n-j+1})\,.
\end{array}
\eq
If $j+k= n+1$ then
\beq{nba1}
\begin{array}{ll}
\cT^1_{(0,0),(j,n-j+1)}=2\si_1\otimes E_{j,n-j+1}\,,&
\cT^0_{(0,1),(j,n-j+1)}=2\si_0\otimes E_{j,n-j+1}\,,\\
\cT^1_{(1,0),(j,n-j+1)}=2\si_2\otimes E_{j,n-j+1}\,,&
\cT^0_{(1,1),(j,n-j+1)}=2\si_3\otimes E_{j,n-j+1}\,.
\end{array}
\eq

It will be convenient to select the diagonal elements of the basis.
Let $e_j=E_{jj}$. Then
\\
 for $n=2l$
\beq{dia}
\begin{array}{c}
  \cH_{(0,a),j}=\cT^0_{(0,a),jj}=\si_0\otimes(e_{j}-(-1)^ae_{n-j+1})\,,\\
\cH_{(1,a),j}=\cT^0_{(1,a),jj}=\si_3\otimes(e_{j}-(-1)^ae_{n-j+1})\,,
\end{array}
~~0\leq j\leq l\,.
\eq
 For $n=2l+1$
\beq{dia1}
\begin{array}{c}
  \cH_{(0,a),j}=\cT^0_{(0,a),jj}=\si_0\otimes(e_{j}-(-1)^ae_{n-j+1}+\de_{a,1}e_{l+1})\,,\\
\cH_{(1,a),j}=\cT^0_{(1,a),jj}=\si_3\otimes(e_{j}-(-1)^ae_{n-j+1}+\de_{a,0}e_{l+1})\,,
\end{array}
~~0\leq j\leq l\,,
\eq
$$
\begin{array}{c}
  \cH_{(0,1),l+1}=\cT^0_{(0,a),l+1,l+1}=\si_0\otimes e_{l+1}\,,\\
\cH_{(1,0),l+1}=\cT^0_{(1,a),l+1,l+1}=\si_3\otimes e_{l+1}\,.
\end{array}
$$


Consider the action of $\nu$ and $\la$ on the basis. From (\ref{an}) we find
\beq{na}
\nu(\cT^\al_{(a,b)})=(-1)^b\cT^\al_{(a,b)}\,,
\eq
 It follows from (\ref{lam}) that the adjoint action of $\La$
takes the form $Ad_\La(\cT^\al_{(a,b)})=Ad_{\si_1\otimes Id_n}(\cT^\al_{(a,b)})$.
Therefore
\beq{laa1}
Ad_\La(\cT^\al_{(a,b)})=(-1)^a\cT^\al_{(a,b)}\,.
\eq

The Killing form $Tr(E_{JK}E_{IM})=\de_{KI}\de_{JM}$ in $\gln$ leads to
the following norm for the generators
\beq{kf1}
\lan\cT^\al_{(a,b),jk},\cT^\be_{(c,d),im}\ran=4\de^{\al\be}\de_{ac}\de_{bd}\de_{ki}\de_{jm}\,.
\eq

 In what follows we need the commutation relations of $\cH_{(0,0),i}$ with other
 generators. For $n=2l+1$ and $j\leq k$
\beq{cra1}
[\cH_{(0,0),i},\cT^\al_{(a,b),jk}]=\left\{
\begin{array}{lll}
1. &(\de_{ij}-\de_{ik})\cT^\al_{(a,b),jk}\,,  & j\,,\,k\leq l\,, \\
2.& \de_{ij}\cT^\al_{(a,b),jk}\,,  & j\leq l\,,\,k=l+1\,, \\
3. &(\de_{ij}+\de_{ik})\cT^\al_{(a,b),jk}\,,  & j\leq l\,,\,k>l+1\,, \\
4.  &\de_{i,n-k+1}\cT^\al_{(a,b),jk}\,, & j= l+1\,,~k\geq l+1\,,\\
5.& (\de_{ij}-\de_{ik})\cT^\al_{(a,b),jk}\,,  & j,k>l+1\,.
\end{array}
\right.
\eq
For $j\geq k$ the right hand side changes sign.
If $n=2l$ then the lines 2. and 4. omit.

\subsubsection{Bundle  $\Ab_{SL}^{\nu,\La}(C_\tq)$}

The bundle $\Ab_{SL}^{\nu,\La}(C_q)$ is defined by the transitions acting on its sections
$\G(\Ab_{SL}^{\nu,\La})$
as
\beq{w1}
\Psi(-t)=\nu(\Psi(t))\,,
\eq
\beq{w2}
\Psi(\tq t)=Ad_{\La}\Psi(t)
\eq
(compare with (\ref{tw1}), (\ref{tw2})).
These conditions are consistent because of (\ref{cc3}). In particular, (\ref{cc3}) means that
 the bundle $\Ab_{SL}^{\nu,\La}(C_q)$ defined by (\ref{w1}), (\ref{w2}) is topologically non-trivial
(see (\ref{cc1})).

 The $\nu$-invariant subgroup of $\SLN$ is the symplectic group Sp$(n,\mC)$. Its subgroup
 $$
 G_0=\{f\in {\rm Sp}(n,\mC)\,|\,Ad_{\La}f=f\}
 $$
 preserves (\ref{cc3}). It plays the role of the constant gauge transformations.
 Consider the diagonal gauge transformations.
Let $\vec u=(u_1,\dots,u_{l})$, where $l=[n/2]$ and $\ti{\vec u}=(u_l,\ldots,u_1)$. Define the diagonal matrix
 $\bfu\in\,\cH_{00}\subset Lie(G_0)$ (\ref{dia}), (\ref{dia1})
\beq{bfu}
\bfu=\left\{
\begin{array}{lc}
  \di(\vec u,-\ti{\vec u},\vec u,-\ti{\vec u})\,, & n=2l\,, \\
  \di(\vec u,0,-\ti{\vec u},\vec u,0,-\ti{\vec u})\,, &  n=2l+1\,.
\end{array}
\right.
\eq
Since $\bfu\in\,\cH_{00}$ we can replace (\ref{w2}) by the more general condition
\beq{w3}
\Psi(\tq t)=Ad_{\bfe(\bfu)\La}\Psi(t)\,.
\eq
It does not break the consistency condition (\ref{cc3}).
The vector $\vec u$ is the tangent vector to the moduli space of the bundle
$Bun(\Ab_{SL}^{\nu,\La}(C_\tq))$. As above, $\vec u$ is defined up to the action of.
 Let $W^\La_0$ is the $\La$-invariant subgroup of the Weyl group
$W($sp$(n,\mC))$, and $P^\vee_0$ is
$\La$-invariant coweight lattice of  sp$(n,\mC)$.
The lattice $P^\vee_0$ has the rank $l=\left[\frac{n}2\right]$.
In terms of the basic vectors in $\gh\subset\sln$ its generators have the form
$$
\varpi^\vee_1=(e_1-e_n+e_{n+1}-e_{2n})\,,\ldots,\varpi^\vee_l=\sum_{k=1}^l(e_k-e_{n-k+1}+
e_{n+k}-e_{2n-k+1})\,.
$$
The Bernstein-Schwarzman group
is defined as
 \beq{BS1}
 W^0_{BS}(r)=W^\La_0\ltimes( \tau P_0^\vee \oplus  P_0^\vee)\,.
\eq
In particular the shifts act on $\bfu$ (\ref{bfu}) as
$$
\bfu\sim\bfu+\ga_1+\ga_2\tau\,,~~\ga_1\,,\ga_2\in P_0^\vee\,.
$$

 In this way the moduli space of
 $\Ab_{SL}^{\nu,\La}(C_\tq)$  is defined as
 \beq{msa}
 \fbox{$
Bun(\Ab_{SL}^{\nu,\La}(C_\tq))\sim\cH_{00}/ W^0_{BS}\,.
$}
\eq

Represent the sections of the bundle $\Ab_{SL}^{\nu,\La}(C_q)$ in the form
\beq{dcs}
\Psi(t)=\sum_{\al,a,b=0,1}\psi^\al_{a,b}(t)+\xi_{a,b}(t)\,.
\eq
\beq{des}
\psi^\al_{a,b}(t)=\sum_{j,k=1}^{l}\psi^\al_{(a,b),jk}(t)\cT^\al_{(ab),jk}\,,~~
\xi_{a,b}(t)=\sum_{j=1}^{l}\xi_{(a,b),j}(t)\cH_{(a,b),j}\,.
\eq

The section takes value in $\sln$ if
\beq{ssl}
\begin{array}{ll}
  \sum_{j=1} ^{l}\xi_{(01),j}(t)=0\,, & {\rm for}\,n=2l\,, \\
  2\sum_{j=1} ^{l}\xi_{(01),j}(t)+\xi_{(01),l+1}(t)=0\,, & {\rm for}\,n=2l+1\,.
\end{array}
\eq

The  quasi-periodicity properties of the sections in this basis follow from (\ref{na}), (\ref{laa1}),
(\ref{cra1})
\beq{qp2a}
\psi^\al_{a,b}(-t)=(-1)^b\psi^\al_{a,b}(t)\,,~~\xi_{a,b}(-t)=(-1)^b\xi_{a,b}(t)\,,
~~\xi_{a,b}(\tq t)=(-1)^a\xi_{a,b}(t)\,,
\eq
\beq{qp3}
\psi^\al_{(a,b),jk}(\tq t)=u_{jk}\psi^\al_{(a,b),jk}(t)\,,
\eq
where
\beq{qp4}
u_{jk}=\left\{
\begin{array}{lll}
1. &(u_j-u_k)\,,  & j\,,\,k\leq l\,, \\
2.& u_{j}\,,  & j\leq l\,,\,k=l+1\,, \\
3. &(u_{j}+u_{k})\,,  & j\leq l\,,\,k>l+1\,, \\
4.  &u_{n-k+1}\,, & j= l+1\,,~k\geq l+1\,,\\
5.& (u_{j}-u_{k})\,,  & j,k>l+1\,.
\end{array}
\right.
\eq


\subsection{Twisted vector bundles}

To  construct vector bundles over $C_\tq$ with non-trivial characteristic classes
we replace the adjoint bundle $\Ab_G$ (\ref{tw1}) by the vector bundle $E_G$ defined as follows.
Consider a (reducible) representation $\pi$ of the Lie algebra $\gg$ (\ref{alge})
in $V$. Let $V_\eta$ be an irreducible $\gg$-module with the highest weight $\eta\in P$.
Assume that the diagram automorphism $\nu$ acts in a such way that $\nu\eta\neq\eta$.
  The module $V$ is a sum the highest modules
\beq{vd}
V=\oplus_{j=0}^{r-1}V_{\nu^j(\eta)}\,,~~\eta\in P\,.
\eq
For example, we can take
$$
\begin{array}{llccc}
\gg & \eta &V_\eta& \nu(\eta)& V_{\nu(\eta)}  \\
\hline
A_{n-1} & \varpi_1 & \underline{n} &\varpi_{n-1}
&\bar{\underline{n}}  \\
 D_n & \varpi_n&{\rm left~ spinors} & \varpi_{n-1}&{\rm right~ spinors}  \\
 E_6 & \varpi_1&\underline{27} & \varpi_6& \bar{\underline{27}}\\
 D_4& \varpi_1 &\underline{8}^V & \varpi_{3}\,,~\nu^2(\eta)=\varpi_{4}&
\underline{8}^L\,,~\underline{8}^R
 \end{array}
$$
\nopagebreak
\begin{center}
\texttt{Table 1.}
Examples of representations corresponding to non-trivial bundles
\end{center}

We define the vector bundle $E_{\bG}$ as the  associated bundle $E_{\bG}=\cE_{\bG}\times_{\bG}V$.
The sections $\psi(t)\in\G(E_{\bG})$ are transformed as (compare with (\ref{tt1}))
\beq{snt}
\psi(t\om)=\nu(\psi(t))\,,~~(\om=\bfe(1/r))\,,
\eq
\beq{snt1}
\psi(\tq t)=R(t)\psi(t)\,,~~R\in\pi(g)\,,~~g\in\bG\,.
\eq
From (\ref{snt}) we find that the action of $\nu $ changes components in the decomposition (\ref{vd}).
If, for example, $\psi(t)$ takes value in $V_{\nu^i(\eta)}$ then
$\psi(t\om)$ takes value in $V_{\nu^{i+1}(\eta)}$.

 The gauge transformations of the
sections $\G(E_{\bG})$ preserving the consistency condition (\ref{cc2})
 have the same form as before (\ref{gtj}). Then, locally (\ref{cc2}) takes the form
  (\ref{cco1}).

It is crucial, that in the contrast with the  adjoint bundles the transition
operators $R$ in the vector bundles act in the modules of the group $\bG$.
It means, in particular, that
in (\ref{cco1}) $\xi$ is defined up to  $\ga\in Q$.

Similar to the lattice  $P^\Upsilon$ (\ref{pip}) define the lattice
\beq{pip1}
Q^\Upsilon= im\,(\Upsilon)\,,~Q^\Upsilon=\{\xi\in Q\,|\,\xi=\nu(\ga)-\ga\,,~\ga\in Q\}\,.
\eq
Consider the r.h.s of (\ref{cco1}). It is an element of $ P^\Upsilon$ (\ref{pip}).
On the other hand, since $\bfx=\bfe(\bfu)$ is an element of the Cartan subgroup
${\bar\cH}\subset\bG$ the element $\bfu\in\gh$ is defined up to elements
from $Q$.
The $Q$-action on $\bfu$ in (\ref{cco1}) means that $\xi$ is defined up to elements
from $Q^\Upsilon$.
In other words, $\xi$ can be considered as an element from the quotient group
$$
\xi\in P^\Upsilon/Q^\Upsilon=\cZ^\nu(\bG)\subseteq\cZ(\bG)\sim P/Q\,,
$$
The condition (\ref{imi}) allows one to describe the quotient explicitly in terms of
the basis of fundamental weights.
In Appendix A we prove that it has the following structure

\newpage

\begin{center}
\begin{tabular}{|c|c|c|c|c|c|c|}
 \hline

&$\gg$ & $\gg_0$ &$\cZ^\nu(\bG)= P^\Upsilon/Q^\Upsilon$ & $\cZ(\bG)=P/Q$ &Weights generating $\cZ^\nu(\bG)$ \\
 \hline
 &1 &2 &3 & 4 &5\\
 \hline
1.& $A_{2n-1}$& $C_n$ & $\mZ_n$ & $\mZ_{2n}$ &$\varpi_1$\\
2.& $A_{2n}$& $B_n$ &$\mZ_{2n+1}$ & $\mZ_{2n+1}$&$\varpi_1$\\
3.& $D_{n+1}$& $B_n$ & $\mZ_2$ & $\mZ_4$, or $\mZ_2\oplus\mZ_2$&$\varpi_n$\\
4.& $D_4$ & $G_2$ &$\mZ_2\oplus\mZ_2$ &$\mZ_2\oplus\mZ_2$&$\varpi_1\,,$~$\varpi_3$ \\
5.& $E_6$ & $F_4$ & $\mZ_3$& $\mZ_3$&$\varpi_1$\\
  \hline
\end{tabular}
\end{center}
\begin{center}
\texttt{Table 2.}
The group $\cZ^\nu(\bG)$
\end{center}
Thus, for the group $\SLN$ ($N=2n$) and $Spin_{2n+2}$  the group $\cZ^\nu(\bG)$ does not coincides
with $\cZ(\bG)$.
\begin {rem}
The group $\cZ^\nu(\bG)$ differs from the center $\cZ(\bG_0)$. For example, in the first
line $C_n\to \bG_0=Sp_n$, $\cZ(Sp_n)=\mZ_2$, while $\cZ^\nu(SL_{2n})=\mZ_n$.
\end{rem}


\bigskip

Now consider the transformations of $\bfu$ preserving  (\ref{cco1}).
The elements $\exp\,2\pi\imath\ga$ for $\ga\in \ti Q^\vee$
are the unity elements in $\bG_0$.
Therefore, $\bfu$ in (\ref{gtr}) is an element of the quotient
\beq{gad1}
\Re e\,\bfu\in \gh_0^\mR/\ti Q^\vee\,,~~(\bfu+\ga\sim\bfu\,,~{\rm for~}\ga\in \ti Q^\vee)\,.
\eq
Thus,
$\Re e\,\bfu$ can be chosen from the fundamental domain in $\gh_0^\mR$ under the action of
the affine Weyl group
$\ti W_{aff}=W_0\ltimes\ti Q^\vee$ (\ref{wap}).

The residual gauge transformations of $\bfu$ is the subgroup
 $L_0(\bar\cH)\subset \cG_{\nu}(C_q, \bG)$ (\ref{gga}) of the equivariant maps
 to $\bar\cH$ (see (\ref{geq}) and (\ref{cl}))
$$
 L_0(\bar\cH)=\{f(t)=t^\ga h(t)\,|\,\ga\in r\ti Q^\vee\,,~h(t)=\sum_kh_kt^k\,,~h_k\in\bar\cH\,,
 ~\nu(h_k)=\om^{-k}h_k\}\,.
$$
Similar to (\ref{gtr1}) the moduli space of the $E_{\bG}$-bundles
is the set of pairs $(\bfu\in \gh_0,\,\xi\in P)$ satisfying (\ref{cco1}), and
invariant with respect to the action
\beq{gtr2}
\cP_{E_{\bG}}\,:\,\left\{
\begin{array}{ll}
  \bfu\to \bfu+\ga_1+\ga_2\tau\,, & \ga_1\,,\ga_2\in \ti Q^\vee\,, \\
  \xi\to\xi+\ga_3 & \ga_3\in Q^\Upsilon\,.
\end{array}
\right.
\eq

 We define the corresponding to the lattice $\ti Q^\vee$  the Bernstein-Schwarzman group
\beq{bs2}
\ti W_{BS}=W_0\ltimes(\tau \ti Q^\vee\oplus \ti Q^\vee)\,.
\eq

Thus, for a fixed $\zeta=\bfe(\xi)\in\cZ^\nu(\bG)$ we can identify
the moduli space with the fundamental domain of the Bernstein-Schwarzman group $\ti W_{BS}$.
\beq{mse}
\fbox{$
 Bun_{\si,\zeta}(E_{\bG}(C_\tq))\sim \gh_0/\ti W_{BS}(r)\,.
$}
\eq


%

Thus, the general solutions of the equation (\ref{cco1}) has the form
\beq{gs}
\bfu+\varpi_j\,,~\bfu\in  Bun_{\si,\zeta}(E_{\bG}(C_q))\,,
\eq
and $\varpi_j$ is a fundamental weight of the Lie algebra $\gg$ satisfying the equation
\beq{gw}
\nu(\varpi_j)-\varpi_j=\xi\,.
\eq
Solutions of this equation are given in  the column 5 of Table 2.
The proof is contained  in Appendix A.

The element $\xi$ defines the characteristic class of the $E_{\bG}$-bundle.
Note, that the moduli space are independent on the characteristic class $\zeta$.

As above to pass to the general automorphism $\si_{r,s}=\nu\bfe(\kappa)$ (\ref{cq1}), (\ref{efs})
we apply the gauge transformation $f=t^{p\kappa}$ (\ref{tka}).

%



\section{Higgs bundles}

\setcounter{equation}{0}

\emph{The Higgs bundle} $\cH(G,\Si_g)$ over the Riemann surface $\Si_{g}$ is the pair $(E_{\bG},\Phi)$,
 where $E_{\bG}=\cE\times_GV$ is
 a holomorphic vector $G$-bundle, associated with the principal bundle $\cE$ and
 $\Phi$ is a Higgs field.
 It is a holomorphic 1-form on $\Si_{g}$ taking values in $End\,E_{\bG}$.
  The Higgs bundle can be equipped with a symplectic form.
 The group of automorphisms of the underlying vector bundle can be lifted
 to the Higgs field in a such  way that it acts as the symplectomorphisms of
the Higgs bundle.

In what follows we describe the Higgs bundles related to
constructed above three types of holomorphic $G$-bundles over elliptic curves
  $\Ab_G$, $\Ab_{SL}^{\nu,\La}$ and $E_{\bG}$ .


\subsection{Higgs bundles related to $\Ab_G$ and $E_{\bG}$ bundles}

The Higgs bundle $\cH(G,C_\tq)$  is described by the pair
$(R(t),\Phi(t))$, where $R(t)$ is the transition operator (\ref{tw2}) and $\Phi$ is a holomorphic one-form
in a neighborhood $\cV$ of the contour $\ga_1=\{|t|=|\tq^\oh|\}$ taking values in the endomorphisms
of the bundles $\Phi=\Phi(t)\frac{dt}t\in\,\Om^{(1,0)}(\cV,End(\gg))$.
 There is a well defined symplectic form on $\cH(G,C_\tq)$
\beq{c1}
\Om=\f1{2\pi\imath}\oint_{\ga_1}
\lan(D (R^{-1}(t)\Phi(t))\wedge DR(t)\ran\,.
\eq

 We assume that $\Phi(t)$ is the equivariant map $\Phi(t)\in\cL_0(\gg,t)$ (\ref{ins}).
\beq{sii}
\nu(\Phi(t)=\Phi(t\om)\,,~(\om=\exp\,\frac{2\pi\imath}r)\,.
\eq
The form (\ref{c1}) is invariant with respect to the
$\cG_{\nu}(G)$ -action (\ref{gtj}) and
\beq{str}
\Phi(t)\to f(\tq t)\Phi(t)f^{-1}(\tq t)\,.
\eq
This symplectic action leads to the moment map
$$
\mu\,:\,\cH(G,C_\tq)\to Lie^*(\cG_{\si}(G))\,,
$$
defined by the functional
$$
\cF=\f1{2\pi\imath}\oint_{\ga_1}\lan\Phi(t)\gx(\tq t)\ran-
\f1{2\pi\imath}\oint_{\ga_1}\lan R^{-1}(t)\Phi(t)R(t)\gx(t)\ran\,,~~\gx(t)\in Lie(\cG_{\si}(G))\,.
$$

As in \cite{BDOZ} we find that the zero moment map implies that $\Phi$ can be extended
holomorphically from the boundary $(\ga_1=\{|t|=|\tq^\oh|\}\,,\,\ga_2=\{|t|=|\tq|^{-\oh}\})$
inside the annulus with at most a single pole at $t=1$ and  such that on $\ga_1$
\beq{mm}
\Phi(\tq^{-1}t)=R^{-1}(t)\Phi(t)R(t)\,.
\eq
Since the gauge group preserves a flag at $t=1$ from (\ref{str}) we find that
 the Higgs field $\Phi(t)$ has a pole at $t=1$
we assume that the elements of Lie algebra of the gauge transformations
\beq{res}
Res\,\Phi(t)|_{t=1}=\bfS\,,
\eq
where $S$ is an element of coadjoint $G$-orbit attached to the point $t=1$.
Details can be found in \cite{BDOZ,LOZ1}.

Acting by the gauge transformations on $R(t)$ (\ref{gtr})
$R(t)\stackrel{f(t)} \rightarrow\bfe(\bfu)$ we transform simultaneously
the Higgs field
$$
\Phi(t)\stackrel{f(t)} \rightarrow L(t)\,.
$$
The transformed Higgs field $L(t)$ plays the role of the Lax operator.
In these terms (\ref{mm}) assumes the form
\beq{mm1}
L(\tq^{-1}t)=\bfe(-\bfu)L(t)\bfe(\bfu)\,.
\eq
Solutions of this equation and (\ref{res}) $L=L(\bfS,\bfu,\bfv,t)$ depends on an additional
parameter $\bfv\in\gh^*_0$.

The reduced phase space can be considered as the cotangent bundle $T^*Bun_{\zeta}(C_\tq,G)$
to the moduli space of holomorphic bundles with the quasi-parabolic structure at $t=1$.
 $(\bfS,\bfv,\bfu)$ play the role of local coordinates in $T^*Bun_{\zeta}(C_\tq,G)$. Here $\bfS$ is an element
 of symplectic quotient of a coadjoint orbit of $G_0$ with respect to the Cartan subgroup action.
 We come to this point below.
The coordinates $(\bfv,\bfu)$ are canonical
\beq{pb1}
\{v_j,u_k\}=\de_{jk}\,.
\eq
and
(see (\ref{mse}))
\beq{dco}
\bfu\in
\left\{
\begin{array}{l}
  Bun_{\si,1}(\Ab_G(C_\tq))\sim
\gh_0/\ti W^{ad}_{BS}  \\
  Bun_{\si,\zeta}(E_{\bG}(C_\tq))\sim
\gh_0/\ti W^{sc}_{BS}
\end{array}
\right.
\eq

Summarizing we define the Lax operator by the conditions
We find from (\ref{sii}) and (\ref{mm})
\beq{lin}
1.\,\nu(L(t))=L(t\om^{-1})\,,~~2.\,L(\tq^{-1}t)=Ad_{\bfe(\bfu)}L(t)\,,~~3.\,Res\,L(t)=\bfS\,.
\eq
We preserve the same notation $\bfS$ for the residue of the Lax operator.

Consider the first condition.
From (\ref{ec}) we have
\beq{lgr}
L(t)=\sum_{m=0}^{r-1}L_m(t)\,,~~(r=2,3)\,,
\eq
\beq{cqp}
L_m\in\gg_m\leftrightarrow\nu(L_m)=\om^mL_m\,,~~L_m(t\om)=\om^{-m}L_m(t)\,.
\eq

To define the Lax operator, satisfying 1. and 2.\,(\ref{lin}) we use the root basis
(\ref{inb}), (\ref{rdi}). In the root basis we have
$$
L_0(t)=L_{0|\gh}(t)\gh_0+\sum_{\ti\al\in\ti R}L_{0,\ti\al}(t)\,,
$$
$$
L_m(t)=L_{m|\gh}(t)\gh_m+\sum_{{\ti\al\in R_s}}L_{m,\ti\al}(t)\,.
$$
We decompose the residue 3.(\ref{lin}) of $L$ in the basis (\ref{tb})
$$
\bfS=\sum_{m=0}^{r-1}S_m\,,~~S_0=\sum_{k=1}^nS_0^k\ge_k^0+\sum_{\ti\al\in\ti R}S_{\ti\al}E_{\ti\al}\,,~~
S_m=\sum_{k=1}^nS^k_m\ge^m_k+\sum_{\ti\al\in R_s}S_m^{\ti\al}\gt^m_{\ti\al}\,.
$$
It is convenient to represent the Cartan part $S_0$  in the  coroot basis (\ref{pid})
\beq{rtc}
S^\gh_0=\sum_{k=1}^nS_0^k\ge_k^0=\sum_{\ti\al\in\Pi}S_0^{\ti\al} H_{\ti\al}\,.
\eq

The invariant component of the Lax operator can be expressed in terms of the functions $g(s,t^r|q)$ (\ref{kro})
 and $E_1(t^r|q)$ (\ref{ef1})
\beq{la1}
L_0(t)=\sum_{k=1}^n(v_k+S_0^k E_1(t^r|q))\ge^0_k+
\sum_{\ti\al\in \ti R}S_{\ti\al}g(s_{\ti\al}^{-1}(\bfu),t^r|q)E_{\ti\al}\,,
~~s_{\ti\al}(\bfu)=\exp \,2\pi\imath\bfu_{\ti\al}\,.
\eq
It satisfies (\ref{lin}) due to (\ref{gf}), (\ref{spe}) and (\ref{res1}).
The Cartan part $L^\gh_0(t)$ can be rewritten in the simple coroot basis (see (\ref{rtc}))
\beq{cla}
L^\gh_0(t)=\sum_{k=1}^n(v_k+S_0^k E_1(t^r|q))\ge^0_k=
\sum_{k=1}^nv_k\ge^0_k+\sum_{\ti\al\in\ti\Pi}S^{\ti\al}_0 E_1(t^r|q))H_{\ti\al}\,.
\eq

The terms $S_0^k E_1(t|q) \ge^0_k$ in (\ref{la1}) violate the residue theorem
and, due to (\ref{gf0}),  brake the symmetry  2.(\ref{lin}).
The origin of this contradiction is that there is residual gauge symmetry.
In fact, these terms vanish after the applying the symplectic reduction with respect to the action
of the invariant Cartan subgroup ${\cal H}_0$. The conditions $S_0^k=0$ play the role of the moment constraints.
(See details in \cite{LOSZ1}).
We neglect this term in calculations of the Hamiltonians.
 The described below $r$-matrix defines the Poisson brackets before the residual reduction.
For this reason it contains the corresponding anomalous term (see (\ref{rm1})).
Note that $L_0$ is the standard Lax operator for the spin Calogero-Moser system related to the invariant
subalgebra $\gg_0$.

The other components of the Lax operator assume the following form.
Introduce $r$ functions
\beq{fj}
g^{(m)}(s,t|q)=t^{m}g(sq^{m/r},t^r|q)\,,~~
(m=0,\ldots,r-1)\,,~~(g^{(0)}(s,t^r|q)=g(s,t^r|q))\,.
\eq
$$
g^{(m)}(s,t|q)=\bfe\left(\frac{mz}r\right)\phi(u+\frac{m\tau}r,z|\tau)\,,~~t=\bfe(z/r)\,,~s=\bfe(u)
\,,~~q=\tq^r\,.
$$
They have the quasi-periodicities
\beq{qp2}
g^{(m)}(s,t\cdot\om|q)=\om^mg^{(m)}(s,t|q)\,,
\eq
\beq{qp5}
s\tq^{m}g^{(m)}(s,t\tq|q)-g^{(m)}(s,t|q)=\de(t,1)-1\,.
\eq
Similarly to (\ref{res1}) we have
\beq{re1}
Res\,g^{(m)}(s,t|q)|_{s=1}=1\,,~~Res\,g^{(m)}(s,t|q)|_{t=1}=1\,.
\eq
In terms of these functions the other components of the Lax operators can be written as
\beq{la2}
L_m(t)=\sum_{k=1}^nS^k_mg^{(-m)}(1,t^r|q)\ge^m_k+
\sum_{\ti\al\in R_s}S_m^{\ti\al}g^{(-m)}(s_{\ti\al}^{-1}(\bfu),t^r|q)\gt^m_{\ti\al}\,.
\eq
Due to  (\ref{qp2}) and (\ref{qp5}) these components have the proper quasi-periodicities (\ref{cqp}).

For the $E_{\bG}$ bundles we should replace $\bfu$ on $\bfu+\varpi_j$, where $\varpi_j$ are solutions of
  (\ref{gw}) taken from Table 2. Then from (\ref{la1}) and (\ref{la2}) we find
\beq{la3}
L_0(t)=\sum_{k=1}^n(v_k+S_0^k E_1(t^r|q))\ge^0_k+
\sum_{\ti\al\in \ti R}S_{\ti\al}g(s_{\ti\al}^{-1}(\bfu+\varpi_j),t^r|q)E_{\ti\al}\,,
\eq
\beq{la4}
L_m(t)=\sum_{k=1}^nS^k_mg^{(-m)}(1,t^r|q)\ge^m_k+
\sum_{\ti\al\in R_s}S_m^{\ti\al}g^{(-m)}(s_{\ti\al}^{-1}(\bfu+\varpi_j),t^r|q)\gt^m_{\ti\al}\,.
\eq

\begin{rem}
In the similar way we can consider meromorphic Higgs field on $C_\tq$ with $n>1$ simple poles.
 In this way, instead of the twisted elliptic
 Calogero-Moser systems, that will be constructed below, we come to the twisted Gaudin
 systems. The quantum non-autonomous analog of the latter system is the Knizhnik-Zamolodchikov-Bernard
 equations. They will be considered in last section.
 \end{rem}


\subsection{Hamiltonians}

The quadratic Hamiltonian $H$ can be found from the expansion of the $\tr L^2(t)$.
It is double-periodic  function on $C_\tq$ with at most second order pole.  Therefore,
\beq{exl}
\oh\tr\,(L(t))^2=H+E_2(t)C_2\,,
\eq
where $C$ is the second Casimir function corresponding to the coadjoint $G$-orbit
$$
C_2=\oh\sum_{m=0}^{r-1}(\bfS,\bfS)\,.
$$
To calculate it we use (\ref{blf}), (\ref{kfo})
$$
\oh\sum_{m=0}^{r-1}\tr\,\left(L_m(t)L_{r-m}(t)\right)=
\left\{
\begin{array}{ll}
  \oh(L_0(t),L_0(t))+\oh (L_1(t)L_1(t)) & r=2\,, \\
  \oh(L_0(t)L_0(t))+(L_1(t)L_2(t)) & r=3\,.
\end{array}
\right.
$$
Then it follows from (\ref{fi1}) that
$$
\oh\sum_{m=0}^{r-1}\left(L_m(t),L_{r-m}(t)\right)=
E_2(t)C_2(\bfS)+H(\bfv,\bfu,\bfS)\,,
$$
\beq{ha}
H(\bfv,\bfu,\bfS)=H_0(\bfv,\bfu,\bfS)+H'(\bfu,\bfS)\,,
\eq
where
$$
H_0(\bfv,\bfu,\bfS)= {\rm const.~part~}\oh(L_0(t),L_0(t))\,,
$$
$$
H'(\bfu,\bfS)=\left\{
\begin{array}{ll}
 {\rm const.~part~} \oh (L_1(t)L_1(t)) & r=2\,, \\
 {\rm const.~part~} \oh(L_1(t)L_2(t)) & r=3\,.
\end{array}
\right.
$$
$$
H_0(\bfv,\bfu,\bfS)=\oh\sum_{k=1}^nv_k^2+
\sum_{\ti\al\in\ti R}S_0^{\ti\al}S_0^{-\ti\al}E_2(\bfu_{\ti\al})\,,
$$
$$
H'(\bfu,\bfS)=\sum_{m\neq 0}\left(-\oh\sum_{k=1}^{n}S_k^m S_k^{r-m}E_2(m\tau/r)+
\sum_{\ti\al\in R_s}S^{\ti\al}_m S^{-\ti\al}_{r-m}E_2(\bfu_{\ti\al}-m\tau/r)\right)\,.
$$

For the $E_{\bG}$-bundles we have
$$
H_0(\bfv,\bfu,\bfS)=\oh\sum_{k=1}^nv_k^2+
\sum_{\ti\al\in\ti R}S_0^{\ti\al}S_0^{-\ti\al}E_2(\bfu_{\ti\al}+\varpi_{j\,\ti\al})\,,
$$
$$
H'(\bfu,\bfS)=\sum_{m\neq 0}\left(-\oh\sum_{k=1}^{n}S_k^m S_k^{r-m}E_2(m\tau/r)+
\sum_{\ti\al\in R_s}S^{\ti\al}_m S^{-\ti\al}_{r-m}E_2(\bfu_{\ti\al}+\varpi_{j\,\ti\al}-m\tau/r)\right)\,.
$$


\subsection{CM system corresponding to $\Ab_{SL}^{\nu,\La}(C_q)$ bundle}

Similarly to (\ref{lin}) the Lax operator is fixed by the conditions
$$
1.\,\nu(L(t))=L(-t)\,,~~2.\,L(q^{-1}t)=Ad_{\bfe(\bfu)\La}L(t)\,,~~3.\,Res_{t=1}\,L(t)=\bfS\,.
$$
Here $\bfS$ is decomposed in the basis (\ref{nba})--(\ref{dia1})
\beq{dsb}
\bfS=\sum_{a,b=0,1}\left(
\sum_{j,k=1}^{l}\sum_{\al=0,1}S^\al_{(a,b),jk}\cT^\al_{(ab),jk}+\sum_{j=1}^{l}S_{(a,b),j}\cH_{(a,b),j}
\right)\,.
\eq

According with (\ref{dcs}) decompose $L(t)$ as
$$
L(t)=\sum_{\al,a,b=0,1}L^\al_{(a,b)}(t)+\sum_{a,b=0,1}L^{diag}_{(a,b)}(t)\,.
$$
Consider the diagonal part of the Lax operator. If $a=b=0$ then
$$
L^{diag}_{(0,0)}(t)=
\sum_{j=1}^l(v_j+S_{(0,0),j}E_1(t^2|q))\cH_{(0,0),j}\,.
$$
Otherwise,
$$
L^{diag}_{(a,b)}(t)=\sum_{j=1}^l SS_{(0,0),j}t^bg(\tq^b\bfe(a/2),t^2|q)\cH_{(a,b),j}\,.
$$
The non-diagonal part is
$$
L^\al_{(a,b)}(t)=\sum_{j,k=1}^{l}L^\al_{(a,b),jk}(t)\cT^\al_{(ab),jk}\,,
$$
where the component satisfy the quasi-periodicity conditions
$$
L^\al_{(a,b),jk}(-t)=(-1)^bL^\al_{(a,b),jk}(t)\,,~~
L^\al_{(a,b),jk}(\tq t)=(-1)^a\bfe(u_{jk})L^\al_{(a,b),jk}(t)\,.
$$
Then we find
$$
L_{(a,b)jk}(t)=S^\al_{(a,b)jk}t^bg(\bfe(-u_{jk}+a/2)\tq^b,t^2|q^2)=S^\al_{(a,b)jk}\bfe(bz/2)
\phi(-u_{jk}+\frac{a+b\tau}2,z|\tau)\,.
$$

%
The Hamiltonian we find from the expansion (\ref{exl}) using the Killing form (\ref{kf1})
and the Fay identity (\ref{fi1}).
As above, we assume that $S_{(0,0),j}=0$ for $j=1,\ldots,l$.
Then the Hamiltonian $H=\f1{8}(const.~part~\tr\,L^2(t)$ takes the form
$$
H=\oh\sum_{j=1}^lv_j^2+\sum_{a+b>0}\sum_{j=1}^lS^2_{(0,0),j}E_2(\frac{a+b\tau}2|\tau)+
\sum_{\al,a,b}\left(\sum^l_{j\neq k,\,j\neq n-k+1 }(S^\al_{(a,b)jk})^2E_2(u_j\pm u_k+\frac{a+b\tau}2|\tau)\right.
$$
\beq{ha1}
\left.
+\sum^l_{j=1}(S^\al_{(a,b)j})^2E_2(u_j+\frac{a+b\tau}2|\tau)
+\sum^l_{j=1}(S^\al_{(a,b)j})^2E_2(2u_j+\frac{a+b\tau}2|\tau)\right)\,.
\eq

\section{Classical r-matrices and KZB equations }

\setcounter{equation}{0}

\subsection{r-matrix}

The classical $r$-matrix is the map $C_q\times C_q\to \gg\otimes\gg$, satisfying the classical
dynamical Yang-Baxter equation (CDYBE) (see below). More exactly,
it is a section of $\gg\otimes\gg$ bundle
   over the moduli space $Bun_{\si,\zeta}(\Ab_G(C_q))$ (\ref{equ})
, or $ Bun_{\si,\zeta}(E_{\bG}(C_\tq))$   (\ref{mse}) defined by
   the properties of the coefficients $g^{(-m)}(\frac{m}r,t^r|q)$ (\ref{qp2}) and (\ref{re1}).

It can be defined the  by means of the Lax operator as
\beq{lr}
r(t)=\frac{\p L_1(t)}{\p S_2}\,,
\eq
Starting with the Lax operator corresponding to the non-trivial characteristic classes
(\ref{la3}), (\ref{la4})  we come to the $r$-matrix
\beq{erm}
r(\bfu,t)=\sum_{m=0}^{r-1}r_m(t)\,,~~(r=2,3)\,,~~r_m(t\om,s)=\om^{-m}r_m(t)\,.
\eq
\beq{rm1}
r_0(\bfu,t)=\sum_{k=1}^n E_1(t^r|q))\ge^0_k\otimes\ge^0_k+
\sum_{\ti\al\in \ti R}|\ti\al|^2g(s_{\ti\al}^{-1}(\bfu+\varpi_j),t^r|q)E_{\ti\al}\otimes E_{\ti\al}\,,
\eq
\beq{rm2}
r_m(\bfu,t)=\sum_{k=1}^ng^{(-m)}(1,t^r|q)(\ge^m_k\otimes \ge^{r-m}_k)+
\f1{r}\sum_{\ti\al\in R_s}|\ti\al|^2g^{(-m)}(s_{\ti\al}^{-1}(\bfu+\varpi_j),t^r|q)(\gt^m_{\ti\al}
\otimes \gt^{r-m}_{-\ti\al})\,,
\eq
where $E_{\ti\al}$ is defined by (\ref{rdi}),
   $g^{(-m)}(\frac{m}r,t^r|q)$ is defined by (\ref{fj}).

   As in \cite{LOSZ2} it can be proved that $r(t)$   satisfies classical dynamical Yang-Baxter
   equation:
 \beq{dfg5}
[r_{12}(t_1,t_2),r_{13}(t_1,t_3)]+[r_{12}(t_1,t_2),r_{23}(t_2,t_3)]+
[r_{13}(t_1,t_3),r_{23}(t_2,t_3)]-
\eq
$$
\sqrt{r}\sum\limits_{m=0}^{r-1}\sum\limits_{\ti\alpha\in\,R_s}\frac{|\ti\al|^2}{2}
\left(
\gt^{\,m}_{\,\ti\al}\otimes \gt^{\,r-m}_{\,-\ti\al}\otimes
\ge_{\al}^{0}\,\p_{\ti\al(\bfu)}g^{(-m)}(s_{\ti\al}^{-1}(\bfu+\varpi_j),(t_1/t_2)^r|q)-
\right.
$$
$$
\left.
\gt^{\,m}_{\,\al}\otimes
\ge_{\al}^{0}\otimes
\gt^{\,r-m}_{\,-\al}\,\p_{\bfu}g^{(-m)}(s_{\ti\al}^{-1}(\bfu+\varpi_j),t_1^r/t_3^r|q)+
\ge_{\al}^{0}\otimes
\gt^{\,m}_{\,\al}\otimes\gt^{\,r-m}_{\,-\al}\,
\p_{\bfu}g^{(-m)}(s_{\ti\al}^{-1}(\bfu+\varpi_j),t_2^r/t_3^r|q)
\right)=0\,.
$$

   It follows from (\ref{gf}) and (\ref{gf0}) that the $r$-matrix  plays the role of
the Green function $G(t/t_1)$ of the equation (\ref{grf}).

The r-matrix  (\ref{erm})-(\ref{rm2}) and described above the Lax operator
 (\ref{la3}), (\ref{la4}) define the Poisson
brackets (\ref{pb1}) via $RLL$-equation:
 \beq{RLLe}
  \left\{ L(t_1)\otimes 1,\,1\otimes L(t_2)
  \right\} \,=\left[L(t_1)\otimes 1+1\otimes L(t_2),
  r(t_1/t_2)\right]-
 \eq
$$
-\frac{\sqrt{r}}{2}\,\sum\limits_{m=0}^{r-1}\sum\limits_{\ti\alpha\in\,R_s}\,
|\ti\alpha|^2 \,\partial_{\bfu}\,g^{(-m)}(s_{\ti\al}^{-1}(\bfu), (t_1/t_2)^r|q)
\,(S^\gh)^{\ti\al}_0\,\gt^{\,m}_{\,\ti\alpha}\otimes
\gt^{r-m}_{-\ti\alpha}\,.
$$
The anomalous last term vanishes upon the moment constraints $(S^\gh)^{\ti\al }_0=0$.


\subsection{KZB equations}

Let $\gM_{l}$ be the space of complex structure of the elliptic curve with $l$ marked points
$(t_1,\ldots, t_l)$.
An open cell in $\gM_{l}$ is described by the coordinates
$(q=\bfe(\tau),t_1,\ldots,t_l)\,, \prod t_j=1)$.
The KZB equations define a flat connection in the bundle of conformal blocks in the WZW theory on $C_q$
over $\gM_{l}$ \cite{Fe,FGK,FW}. In our case the variables in the WZW theory are defined by the quasi-periodic conditions
(\ref{tw1}), (\ref{tw2}). Acting as in \cite{LOSZ3} we come to the following description of
the KZB equations.

Consider  the following differential operators:
  \beq{t2}
 \nabla_a=t_a\p_{t_a}+\hat{\p}^a+\sum\limits_{c\neq a}r^{ac}\,,
  \eq
  \beq{t202}
\nabla_\tau=2 \pi i \p_\tau+\Delta+\frac{1}{2}\sum\limits_{b,d}f^{bd}\,,
 \eq
with
 $$
r^{ac}=\sum\limits_{m=1}^{r-1} \sum\limits_{\ti\alpha \in \ti R}|\alpha|^2
g^{(-m)}(s_{\ti\al}^{-1}(\bfu+\varpi_j),(t_a/t_c)^r) \gt^{m,a}_{\ti\al}\otimes
\gt^{-m, c}_{-\ti\al} +
 \sum\limits_{m=0}^{r-1}\sum\limits_{k=1}^n  g^{(-m)}(1,(t_a/t_c)^r)
 \ge_{k}^{m,a}\otimes\ge_{k}^{-m,c}\,,
$$
$$
f^{ac}=\sum\limits_{m=1}^{r-1} \sum\limits_{\ti\alpha \in \ti R}|\alpha|^2
\p_{\ti\al(\bfu)}g^{(-m)}(s_{\ti\al}^{-1}(\bfu+\varpi_j),(t_a/t_c)^r)
 \gt^{m,a}_{\ti\al}\otimes \gt^{-m, c}_{-\ti\al} +
 $$
 $$
\sum\limits_{k=1}^n \sum\limits_{m=1}^{r-1}  g^{(-m)}(1,(t_a/t_c)^r)
 \ge_{k}^{m,a}\otimes\ge_{k}^{-m,c}+
\oh(E^2_1(t^r_a/t_c^r)-\wp(t^r_a/t_c^r))
 \ge_{k}^{0,a}\otimes\ge_{k}^{0,c}\,,
$$
where $\gt^{k,a}_{\alpha}=1\otimes...1\otimes \gt^{k}_{\alpha}\otimes 1...\otimes 1$ (with $\gt^{k}_{\alpha}$ on the a-th
place) and similarly
 for the generators
 $\ge_{k}^{m,a}$
 \footnote{For brevity we write $\gt^{k,a}_{\alpha}$\,, $\ge_{k}^{m,a} $
 instead of representations of these generators
  in the spaces $V_{\mu_a}$.} and the Weierstrass function $\wp$ is defined in (\ref{a101}).
The following short notations are used here:
 $$
 \hat{\p}^a=r \sum\limits_{\ti\alpha \in \ti\Pi }  \ge_{\ti\alpha} \partial_{\ti{\alpha}}\,,~~
 \Delta=\frac{r}{2}\sum\limits_{\ti\alpha\in \ti\Pi}
   \p^2_{u_ (\ti\alpha)}\,,~~g^{(0)}(1,(t_a/t_c)^r)=E_1((t_a/t_c)^r)\,.
  $$
As in \cite{LOSZ3} it can be proved that the connection (\ref{t2}), (\ref{t202}) is flat.


\section{Appendix A: Structure of Table 2}
\setcounter{equation}{0}
\def\theequation{A.\arabic{equation}}

The group $\cZ(\bG)=P/Q$ can be defined in terms of the Cartan matrix $a_{jk}=(\al_j,\al_k)$.
Since the fundamental weights $\varpi_j$, generating the weight lattice $P$,
 form a basis in $\gh^*$
dual to the basis of simple roots $(\al_j,\varpi_k)=\de_{jk}$, we have the following relations
between two bases in the lattices $P$ and $Q$ $\varpi_k=(a)^{-1}_{kj}\al_j$. In particular,
$ord\,(\cZ(\bG))=\det\,a$.

It follows from (\ref{pip}) that the fundamental weights $\varpi_j$ form the
basis
\beq{pb}
\{\eta_j=\si(\varpi_j)-\varpi_j\}\,,
\eq
 in $P^\Upsilon$, while elements of the root lattice $Q(\gg)$ of the form
 \beq{bq}
 \{\de_j=\si(\al_j)-\al_j\,|\,\al_j\in\Pi(\gg)\}
 \eq
 form a bases in $Q^\Upsilon$. By comparison of these two bases we find the structure of
 quotient group $\cZ^\si(\bG)$.

We construct Table 2 line by line.

\underline{1) $\SLN$}, \\
We use the canonical basis $(e_1,\ldots, e_N)$. Since the fundamental weights of $A_{N-1}$
are
$$
\begin{array}{l}
 \varpi_1=(\frac{N-1}N,-\f1{N},\ldots,-\f1{N})\,,\\
 \varpi_2=(\frac{N-2}N,\frac{N-2}N,\ldots,-\frac{2}{N})\,,   \\
 \ldots \ldots\\
 \varpi_j=(\frac{N-j}N,\ldots\frac{N-j}N,-\frac{j}{N},\ldots,-\frac{j}{N})\,,\\
 \ldots \ldots\\
   \varpi_{N-1}=(\frac{1}N,\frac{1}N,\ldots,\frac{1-N}{N}) \,. \\
\end{array}
$$
we find generators of $P^\Upsilon(A_{N-1})$
\beq{eta}
\eta_j=-\left(\underbrace{1-\frac{2j}N,\ldots, 1-\frac{2j}N,}_j
\underbrace{-\frac{2j}N,\ldots,-\frac{2j}N,}_{N-2j}
\underbrace{1-\frac{2j}N,\ldots, 1-\frac{2j}N}_j\right)\,,
\eq
where
$$
j=\left\{
\begin{array}{ll}
  1,\dots,n-1 & \mbox{for~} N=2n\,, \\
  1,\ldots, n &  \mbox{for~} N=2n+1\,.
\end{array}
\right.
$$

On the other hand, since $\Pi(A_{N-1})=\{e_j-e_{j+1}\,|\,j=1,\ldots,N-1\}$
 we obtain from (\ref{bq})
 \beq{de}
\de_j=\left\{
\begin{array}{lll}
  (0,\ldots,-1,1,0,\dots,0,1,-1,0,\ldots,0) & j=1,\ldots n-1\,, &N=2n,2n+1\,, \\
  (0,\ldots,-1,2,-1,\dots,0) & j=n\,, & N=2n+1\,.
\end{array}
\right.
\eq
It follows from (\ref{eta}) and (\ref{de}) the quotient group
$\mZ^\si(\SLN)=P^\Upsilon(A_{N-1})/Q^\Upsilon(A_{N-1})$ is generated by the element $\eta_1$.
For $N=2n$ it
has the order $n$, while for the odd $N$ the quotient has the order $N$. In this way we have
derived the first two lines in the Table.

\underline{2. SO$(2n+2,\mC)$, ($D_{n+1}$)}\\
In the Cartan subalgebra  $\gh(D_{n+1}$  we choose a canonical bases
$(e_1,\ldots,e_{n+1})$ and the system of simple roots
\beq{ald}
\Pi(D_{n+1})=\{\al_j=e_j-e_{j+1}\,,~j=1,\dots,n\,,~\al_{n+1}=e_n+e_{n+1}\}\,.
\eq
The dual system of the fundamental weights takes the form
\beq{fwd}
\varpi_j=e_1+e_2+\ldots+e_j\,,~j=1,\ldots,n-1\,,
\eq
$$
\varpi_{n}=\oh(e_1+\ldots+e_{n}-e_{n+1})\,,~~
\varpi_{n+1}=\oh(e_1+\ldots+e_{n}+e_{n+1})\,.
$$
The outer automorphism acts as
$$
\si(\al_j)=\left\{
\begin{array}{cc}
  \al_j & 1\leq j<n\,, \\
  \al_n\leftrightarrow\al_{n+1}\,, &
\end{array}
\right.
~~
\si(\varpi_j)=\left\{
\begin{array}{cc}
  \varpi_j &  1\leq j<n\,, \\
  \varpi_n\leftrightarrow\varpi_{n+1} \,.&
\end{array}
\right.
$$
Following (\ref{pb}), (\ref{bq}) we find generators of $P^\Upsilon(D_{n=1})$ and
$Q^\Upsilon(D_{n=1})$
\beq{ded}
\eta_n=\si(\varpi_n)-\varpi_{n}=e_{n+1}\,,~~\de=2 e_{n+1}\,.
\eq
Thus, the quotient group $\mZ^\si(Spin_{2n+2})\sim\mZ_2$ is generated by $\eta$ (\ref{ded}).

\underline{3. $Spin_8,~(D_4)$}\\
Here we have the outer automorphism of order 3.
In terms of notations (\ref{ald}) and (\ref{fwd}) for $n=3$ it acts as
$$
\si(\al_1)=\al_3\,,~\si(\al_2)=\al_2\,,~\si(\al_3)=\al_4\,,~\si(\al_4)=\al_1\,,
$$
$$
\si(\varpi_1)=\varpi_3\,,~\si(\varpi_2)=\varpi_2\,,~\si(\varpi_3)=\varpi_4\,,~
\si(\varpi_4)=\varpi_1\,,
$$
Then $P^\Upsilon$ is generated by the weights
\beq{ed3}
\eta_1=\varpi_1-\varpi_3=\oh(1,-1,-1,1)\,,~~\eta_2=\varpi_3-\varpi_4=-(0,0,0,1)\,,
\eq
while $Q^\Upsilon$ is generated by the elements
$$
\de_1=(1,-1,-1,1)\,,~~\de_2=-2(0,0,0,1)
$$
Then $\eta_1$ and $\eta_2$ are generators of the two groups
$\mZ^\si(Spin_{4})= P^\Upsilon(D_4)/Q^\Upsilon(D_4)\sim\mZ_2\oplus\mZ_2$.

\underline{4. $E_6$}\\
The outer automorphism of  the Lie algebra $E_6$
corresponds to the symmetry of the simple roots $\Pi(E_6)=(\al_j)$
$$
\si\,:\,\al_1\leftrightarrow\al_5\,,~~\al_2\leftrightarrow\al_4\,,~~
\al_{3,6}\leftrightarrow\al_{3,6}\,,
$$
The symmetry of  the fundamental weights
$$
\begin{array}{l}
  \varpi_1 =
  \f1{3}(4\al_1+5\al_2+6\al_3+4\al_4+2\al_5+3\al_6),,\\
   \varpi_2=
   \f1{3}(5\al_1+10\al_2+12\al_3+8\al_4+4\al_5+6\al_6)\,,  \\
     \varpi_3=
     2\al_1+4\al_2+6\al_3+4\al_4+2\al_5+3\al_6 \,,\\
 \varpi_4=
 \f1{3}(4\al_1+8\al_2+12\al_3+10\al_4+5\al_5+6\al_6)\,,  \\
  \varpi_5=
  \f1{3}(2\al_1+4\al_2+6\al_3+5\al_4+4\al_5+3\al_6)\,,  \\
  \varpi_6
  =\al_1+2\al_2+3\al_3+2\al_4+\al_5+2\al_6 \,.
\end{array}
$$
has the similar form
$$
\si\,:\,\varpi_1\leftrightarrow\varpi_5\,,~~\varpi_2\leftrightarrow\varpi_4\,,~~
\varpi_{3,6}\leftrightarrow\varpi_{3,6}\,.
$$
The lattice $P^\Upsilon(E_6)$ is generated by the two elements of weight lattice $P(E_6)$
\beq{ee6}
\eta_1=\varpi_1-\varpi_5=\f1{3}(2\al_1+\al_2-\al_4-2\al_5)\,,~~
\eta_2=\varpi_2-\varpi_4=\f1{3}(\al_1+2\al_2-2\al_4-\al_5)\,.
\eq
and $Q^\Upsilon(E_6)$ by the two elements of the root lattice $Q(E_6)$
$$
\de_1=\al_1-\al_5\,,~~\de_2=\al_2-\al_4\,.
$$
The quotient group $\mZ^\si(E_6)= P^\Upsilon(E_6)/Q^\Upsilon(E_6)\sim\mZ_3$ is
generated by one of the elements $\eta_{1,2}$ (\ref{ee6}).


\section{Appendix B: Some elliptic functions \cite{B,We}}
\setcounter{equation}{0}
\def\theequation{B.\arabic{equation}}
The main ingredient  is the Kronecker function on $s,t\in\mC^*$
\beq{kro}
g(s,t|q)=-\sum_{m\in\mZ}\frac{t^m}{1-q^ms}\,,~~{\rm for~}1>|t|>|q|\,.
\eq
If in addition $1>|s|>|q|$ then
\beq{kr1}
g(s,t|q)=1-\f1{1-t}-\f1{1-s}+\sum_{i,n< 0}s^iq^{in}t^n
-\sum_{i,n> 0}s^iq^{in}t^n\,.
\eq
Then
\beq{res1}
Res\,g(s,t|\tq)|_{s=1}=1\,,~~Res\,g(s,t|q)|_{t=1}=1\,.
\eq
Rewrite (\ref{kr1}) as
\beq{kr}
g(s,t|q)=1+g^+(t)+g^-(t)\,,~~({\rm for~}1>|s|>|q|\,,~1>|t|>|q|)\,,
\eq
$$
g^+(s,t)=-\sum_{i,n\geq 0}s^iq^{in}t^n\,,~~
g^-(s,t)=\sum_{i,n< 0}s^iq^{in}t^n\,.
$$

Consider the distribution on the space of the Laurent polynomials\\
 $\mC[t,t^{-1}]=\{\psi(t)=\sum_lc_lt^n\}$, defined by the functional $\de(t,t_1)$
\beq{de1}
Res_{t_1=0}\,(\de(t,t_1)\psi(t_1))=\psi(t)\,.
\eq
It is represented as formal power series
$$
\de(t_1,t_2)=\sum_{k\in\mZ}t_1^kt_2^{-k-1}=\de_+(t_1,t_2)+\de_-(t_1,t_2)\,,
$$
$$
\de_+(t_1,t_2)=\f1{t_2}\sum_{n\geq 0}(t_1/t_2)^n=\f1{t_1-t_2}~{\rm for~} |t_2|<|t_1|\,,
$$
$$
\de_-(t_1,t_2)=\f1{t_2}\sum_{n> 0}(t_2/t_1)^n=
-\f1{t_1-t_2}~{\rm for~} |t_2|>|t_1|\,.
$$
Fix $t\in\mC^*$. Define two contours in $\mC^*$ with the opposite   orientation:
$$
\ga_1=\{|s|=|t|(1+\ve)\,,~~\ga_2=\{|s|=|t|(1-\ve)\,,~~\ve>0\,,
$$
 and let  $\ga=\ga_1\cup\ga_2$. Then in accordance with (\ref{de1})
\beq{def}
\psi(t)=\f1{2\pi\imath}\oint_{\ga}\psi(t_1)\de(t,t_1)\frac{dt_1}{t_1}\,.
\eq

It follows from (\ref{kro}) that the Kronecker function  plays the role of the Green function
for the difference operator
\beq{gf}
sg(s,q t|q)-g(s,t|q)=\de(t,1)-1\,.
\eq
Let $\psi(t)$ be holomorphic function in a neighborhood of the contour $\ga$, such that
 $Res_{t=1}\frac{\psi(t)}t=0$. Then there exists a holomorphic in $\mC^*$ solution
 $x(t)$ of the difference equation
\beq{dif}
sx(q t)-x(t)=\psi(t)\,.
\eq
It is defined by the integral
\beq{inr1}
x(t)=\f1{2\pi\imath}\oint_{\ga_1}\psi(t_1)g(s,t/t_1|q)dt_1/t_1\,.
\eq

In terms of the theta function the Kronecker function can be represented as
$$
g(s,t|q)=\frac{X(ts|q)P^2(q)}{X(t|q)X(s|q)}\,,~~P(q)=\prod_{n>0}(1-q^n)\,,
$$
where
$$
X(t|q)=4\imath q^{1/8}(t^\oh-t^{-\oh})\prod_{n\geq 1}(1-q^nt)(1-q^nt^{-1})(1-q^n)
=\imath\sum_{n\in\mZ}(-1)^nq^{\oh(n-\oh)^2}t^{(n-1)/2}\,.
$$
In the variables $(u,z)$, $(s=\bfe(u),t=\bfe(z)$) $X(t|q)$ is the odd theta-function\\
$X(t|q)|_{t=\bfe(z)}=\theta(z|\tau)$, and
$$
g(s,t|q)|_{s=\bfe(u),t=\bfe(z)}=\phi(u,z|\tau)=\frac{\theta(u+z)\theta'(0)}{\theta(u)\theta(z)}\,.
$$




The first Eisenstein series are defined as
 $$
 E_1(z|\tau)=\p_z\log\theta(z|\tau)\,.
 $$
 It has a simple poles at the lattice
 \beq{spe}
 Res\, E_1(z|\tau)|_{z=m+n\tau}=1\,.
 \eq
 In the  variable $t=\bfe(z)$ it can be defined on the annulus $1>|t|>|q|$ as the series
 \beq{ef1}
 E_1(t|q)=\pi\imath+2\pi\imath\sum_{m\neq 0}\frac{t^m}{q^m-1}\,.
 \eq
 The function $-\f1{2\pi\imath} E_1(t|q)$ plays the role of the Green function for the
 difference operator on $C_q$
 \beq{gf0}
 -\f1{2\pi\imath} E_1(tq|q)+\f1{2\pi\imath}E_1(t|q)=\de(t,1)-1\,,
\eq
where $\de(t,1)$ is (\ref{def}).
It allows one to continue $E_1(t|q)$ from $C_q$ on $\mC^*$.

Let $E_2(t|q)$ be the second Eisenstein series:
\beq{E2}
E_2(t|q)=4\pi^2\sum_n\frac{q^nt}{(1-q^nt)^2}=-\p_z^2\log\te(z|q)\,.
\eq
 \beq{A.2}
E_2(z|\tau)=-\p_zE_1(z|\tau)=
\p_z^2\log\vth(z|\tau)\,.
\eq
Let $\eta(q)=q^{\frac{1}{24}}\prod_{n>0}(1-q^n)$, and
$$
 \eta_1(q)=\frac{24}{2\pi i}\frac{\p_\tau\eta(q)}{\eta(q)}\,.
 $$
 Then the Weierstrass function is defined as
\beq{a101}
\wp(q,\tau)=E_2(q,\tau)-2\eta_1(q)\,.
\eq

The function $g$ satisfies the Fay identities:
\beq{fi1}
g(s,t|q)g(s^{-1},t|q)=E_2\left(\f1{ 2\pi\imath}\ln t|q\right)-
E_2\left(\f1{ 2\pi\imath}\ln s|q\right)\,,
\eq
\beq{fi2}
g(s_1,t|q)s_2\p_{s_2}g(s_2,t|q)-g(s_2,t|q)s_1\p_{s_1}g(s_1,t|q)=
\eq
$$
\frac{1}{2\pi\imath}
\left(E_2\left(\f1{ 2\pi\imath}\ln s_2|q\right)-E_2\left(\f1{ 2\pi\imath}\ln s_1|q\right)\right)
g(s_1s_2,t|q)\,.
$$



\section{Appendix C: Outer automorphisms of simple algebras \cite{Ka,Ka1,VO}}\label{gi}
\setcounter{equation}{0}
\def\theequation{C.\arabic{equation}}

Let $R$ be the root system of the Lie algebra $\gg$ of rank $l$, $W(R)$ is the Weyl group
 and $Aut(R)\supseteq W(R)$ is the group
of the automorphisms of $R$.
The group $Out(\gg)=\{\nu\}$  is isomorphic to the quotient group $Aut(\Pi)\sim Aut(R)/W(R)$
(see (\ref{sdd})).
This group is non-trivial for the simply-laced algebras
\beq{alge}
\begin{array}{lc}
 R= A_l\,,\,D_l\,,\,E_6 & Aut(R)/W(R)\sim\mZ_2 \,,\\
 R= D_4 & Aut(R)/W(R)\sim\mZ_3 \,.
\end{array}
\eq

The outer automorphisms provide the decompositions of $\gg$ on its eigenspaces
\beq{ec}
\gg=\oplus_{m=0}^{r-1}\,\gg_m\,,~~\nu(\gg_m)=\om^m\gg_m\,,~~(\om=\exp\,(\frac{2\pi\imath}r))\,,~~
r=2\,,~(\,{\rm or}~r=3~\mbox{for }D_4)\,.
\eq
The subspaces $\gg_0$ is an invariant subalgebra of $\gg$ and $\gg_j$ are irreducible
modules of $\gg_0$.

Let $\gh$ be the Cartan subalgebra, $E_\al$ are the root subspaces and
$$
\gg=\gh+\sum_{\al\in R}\mC E_\al
$$
is the root decomposition of $\gg$.
The action of the group $Aut(\gg,\gh,\Pi)$ (\ref{sdd}) preserves this decomposition.
 Then we define the action of $\nu$ on $R$ as
$\nu\to \bar\nu\al$, $\bar\nu\in Aut(\Pi)$
\beq{oa1}
\nu E_\al=E_{\bar \nu\al}\,,~(\al\in R=R^\vee)\,.
\footnote{Since the algebras are simply-laced their root and weight systems coincide with
the coroots and coweight systems. The out-invariant subalgebras are not simply-laced and these
systems are different.}\,.
\eq

Represent the Cartan subalgebra as $\gh=\gh^\mR+\imath\gh^\mR$ and
let
\beq{pi}
\Pi=(\al_1,\ldots,\al_l)\,,~~(l-\mbox{rank }\gg)
\eq
 be a system of  simple roots generating $R$.
Define the Weyl chamber in  $\gh^\mR$
\beq{wc}
C^+=\{x\in\gh^\mR\,|\,\lan x,\al\ran>~\mbox{for }\al\in\Pi\}\,.
\eq

Define the invariant basis in the homogeneous component $\gg_m$ of $\gg$.
 Under the action of $\bar\nu$ the set of roots $R$ is splitted   on the orbits
 on the lengths 1 (invariant subset),
$O_2$ of the length 2 and $O_3$ of the length 3 (for $D_4$).
For all $\gg$ with non-trivial outer automorphisms except sl$(2n+1)$ ($A_{2n}$) define
two subsets of roots
\beq{inr}
\begin{array}{l}
  R_l=\{\ti\al=\al\in R\,|\,\bar \nu(\al)=\al\}\,, \\
  R_s=\{\ti\al=\f1{r}\sum_{m=0}^{r-1}\bar \nu^m(\al)\,|\,\al\in R\,,\,\bar \nu(\al)\neq\al\}\,.
\end{array}
\eq

The subset $\ti R=R_l\cup R_s$ is  the set of roots of the invariant subalgebra $\gg_0$
with respect to the Cartan subalgebra $\gh_0\subset\gg_0$. The system of simple roots
\beq{isr}
\ti\Pi=\{\ti\al_j\,,~j=1,\ldots,n\}
\eq
of the invariant algebra $\gg_0$ is the basis of $\ti R$.

Let $\ti\al_0$ be the highest weight of the
$\gg_1$ module.  It is decomposed  in the root basis as
\beq{dea}
\ti\al_0=\sum_{j=1}^na_j\ti\al_j\,,
\eq
where $n$ is a rank of $\gg_0$.
The extended root system
\beq{edg}
\ti\Pi^{ext}=\ti\Pi\cup(-\ti\al_0)
\eq
defines twisted affine Dynkin diagram.

Let
\beq{bal}
\gt^m_{\ti\al}=\sum_{j=0}^{r-1}\om^{jm}E_{\bar\nu^j\al}\,,~(\ti\al\in R_s)\,,
~\nu(\gt^m_{\ti\al})=\om^m\gt^m_{\ti\al}\,.
\eq
\beq{crd}
\gh=\sum_{m=0}^{r-1}\gh_m\,,~~\nu(\gh_m)=\om^m\gh_m\,.
\eq
Then the subspace $\gg_m$ in (\ref{ec}) has the root decomposition
\beq{inb}
\gg_m=\gh_m+\sum_{\ti\al\in R_s}\mC \gt^m_{\ti\al}\,, ~~m\neq 0\,.
\eq
\beq{rdi}
\gg_0=\gh_0+\sum_{\ti\al\in\ti R}\mC E_{\ti\al}\,,~~E_{\ti\al}=\left\{
\begin{array}{cc}
  E_\al & \al=\ti\al\in R_l\,,\\
  t^0_{\ti\al} & \ti\al\in R_s\,.
\end{array} \right.
\eq
In these terms the set of the root subspaces of $\gg$ is
$$
 E_{\ti\al}\,,\,(\ti\al\in R_l)\,,~~ t^0_{\ti\al}\,\,(\ti\al\in R_s)\,,~~
 \gt^m_{\ti\al}\,,\,(\ti\al\in R_s)\,.
$$

 The   invariant subalgebra $\gg_0$ has the rank
$$
rank\,\gg_0=n\,,~{\rm for~}A_{2n}\,,A_{2n-1}, D_{n+1}
$$
$$
rank\,\gg_0=4\,,~{\rm for~}E_6\,,~~rank\,\gg_0=2\,,~{\rm for~}D_4\,.
$$


Let $(e_1,\ldots,e_l)$ be a canonical basis in $\gh$. The basis in $\gh_m$
for $m\neq 0$ assumes the form $\gh_m=\lan\ge^m_1,\ldots,\ge^m_n\ran\,,$
\beq{inv1}
\ge^m_k=\f1{\sqrt{r}}\sum_{j=0}^{r-1}\om^{jm}e_{\nu^jk}\,\,{\rm for\,}\nu(e_k)\neq e_k\,,
~\nu(\ge^m_k)=\om^{m}\ge^m_k\,,~~(k=1,\dots,n)\,,
\eq
where $n$ is equal to the number of orbits of the $\nu$ action on $\gh$.
In particular, the basis in $\gh_0$ is
\beq{inv5}
\ge^0_k=\f1{\sqrt{r}}\sum_{j=0}^{r-1}e_{\nu^jk}\,\,{\rm for\,}\nu(e_k)\neq e_k\,,
~{\rm and~} \ge^0_i=e_i~{\rm for~} \nu(\ge_i)=\ge_i\,.
\eq


Summarizing, the following set of the  generators $T_a$ $(a=(\ti\al,m,k))$ defines the basis  in $\gg$
\beq{tb}
T_a=\left\{
\begin{array}{cc}
\gg_0 &  \ge^0_k\,(\ref{inv5})\,,~E_{\ti\al}\,(\ref{rdi})\,, \\
  \gg_m\,,~(m\neq 0) &\ge^m_k\,(\ref{inv1})\,,~ \gt^m_{\ti\al}\,(\ref{bal})\,,
\end{array}
m=0,\ldots,r-1\,,~k=1\ldots,n\right\}\,.
\eq

The Killing form on these generators takes the form
\beq{blf}
\begin{array}{ll}
  (\gt^m_{\ti\al},\gt^n_{\ti\be})= r\de^{m+n,r}\de_{\al,-\be}\,, & m=0,\ldots,r-1\,, \\
  (E_{\ti\al},E_{\ti\be})=\de_{\ti\al,-\ti\be}\,, & \ti\al\in R_l\,, \\
  (\ge^m_k,\ge^n_j)=\de^{m+n,r}\de_{j,k}\,. &
\end{array}
\eq

For $\gg=$sl$(2n+1)$ ($A_{2n}$) the invariant subalgebra $\gg_0=B_n$.
$$
\gg=\gg_0\oplus\gg_1\,,~~\gg_0=\mbox{so}(2n+1)\,,
$$
The outer automorphism in the fundamental representation of the algebra  sl$(2n+1)$
can be defined as
\beq{jf}
\nu(x)= -Jx^TJ\,,~~J=\left(
                       \begin{array}{ccc}
                         0 & 0 & j \\
                         0 & 1 & 0 \\
                         -j & 0 & 0 \\
                       \end{array}
                     \right)\,,~~
                     j=(E_{k,n-k+1})\,.
\eq
The anti-invariant part has the form
$$
\gg_1=\left(
        \begin{array}{ccc}
          A & \ti X^T & B \\
          Y & 0 & \ti X \\
          C & \ti Y^T & \ti A^T \\
        \end{array}
      \right),~
 \ti A=jAj\,,~\ti X=Xj\,,~\ti Y=Yj\,,~B=-\ti B^T\,,~C=-\ti C^T\,.
$$
The root system of $\gg_0$ is $\ti R_{B_n}=R_l\cup R_s$, where the short and long roots are
\beq{rao}
\begin{array}{l}
  R_l=\{\ti\al=\oh(\al+\nu(\al))\,|\,\al\neq\nu(\al)~\mbox{and}~(\al,\nu(\al))=0\}\,, \\
  R_s=\{\ti\al=\oh(\al+\nu(\al))\,|\,\al\neq\nu(\al)~\mbox{and}~(\al,\nu(\al))\neq 0\}\,.
\end{array}
\eq
Let $ht\,(\al)$ be the height of $\al\in R$. The root spaces of $\gg_0$ are
\beq{rso}
E_{\ti\al}=\left\{
\begin{array}{ll}
  E_\al-(-1)^{ht\,(\al)}E_{\nu(\al)} & \al\in R_l\,, \\
  \sqrt{2}(E_\al-(-1)^{ht\,(\al)}E_{\nu(\al)}) & \al\in R_s\, .
\end{array}
\right.
\eq
The other root subspaces are
\beq{ors}
\gt^1_{\ti\al}=\sum_{j=0}^{r-1}(-1)^{j}E_{\nu^j\al}\,.
\eq

In the representation (\ref{jf})
\beq{bjf}
\begin{array}{l}
  R_l=\{\ti\al=\oh(e_j\pm e_k-(e_{2n+2-j}\pm e_{2n+2-k}))\,,~j=1\ldots,n\}\,, \\
  R_s=\{\ti\al=\oh(e_j-e_{2n+1-j})\,,~j=1\ldots,n\}\,.
\end{array}
\eq
\beq{bjfr}
E_{\ti\al}=\left\{
\begin{array}{ll}
  E_{jk}-E_{2n+2-k,2n+2-j}&\ti\al\in R_l\,, \\
   E_{j,n+k}-E_{2n+2-k,2n+2-j} &\ti\al\in R_l\,, \\
    E_{n+k,j}-E_{2n+2-j,2n+2-k} &\ti\al\in R_l\,, \\
  \sqrt{2}(E_{j,n+1}-E_{n+1,2n+2-j}) &\ti\al\in R_s\,,\\
  \sqrt{2}(E_{n+1,j}-E_{2n+2-j,n+1}) &\ti\al\in R_s\,
\end{array}
\right.
\eq
\beq{bjfr1}
\gt^1_{\ti\al}=\left\{
\begin{array}{ll}
  E_{jk}+E_{2n+2-k,2n+2-j}
  &\mbox{long roots}\,,
  \\
   E_{j,n+k}+E_{2n+2-k,2n+2-j}
   &\mbox{long roots}\,,
   \\
    E_{n+k,j}+E_{2n+2-j,2n+2-k}
    &\mbox{long roots}\,,
    \\
  \sqrt{2}(E_{j,n+1}+E_{n+1,2n+2-j})
 &\mbox{short roots}\,,
  \\
  \sqrt{2}(E_{n+1,j}+E_{2n+2-j,n+1})
  &\mbox{short roots}\,.
\end{array}
\right.
\eq

The twisted canonical basis  in $\gh\subset$sl$(2n+1)$ is
\beq{ocb}
\ge_k^0=\f1{\sqrt{2}}(e_k-e_{2n+1-k})\,,~~\ge_k^1=\f1{\sqrt{2}}(e_k+e_{2n+1-k})
\,,~~(k=1,\dots,n)\,.
\eq

The expressions (\ref{bjfr})-(\ref{bjfr1}) define the basis in $\gg=$sl$(2n+1)$
with the Killing form
\beq{kfo}
\begin{array}{ll}
(E_{\ti\al},E_{\ti\be})=4\de_{\ti\al,-\ti\be}\,,   & \ti\al\in R_l   \,,        \\
(E_{\ti\al},E_{\ti\be})=8\de_{\ti\al,-\ti\be}\,,   &\ti\al\in R_s\,,\\
(\gt^1_{\ti\al},\gt^1_{\ti\be})=4\de_{\ti\al,-\ti\be}\,,   & \mbox{long roots}\,,\\
(\gt^1_{\ti\al},\gt^1_{\ti\be})=8\de_{\ti\al,-\ti\be}\,,   & \mbox{short roots}\,,\\
(\ge^m_k,\ge^n_j)=4\de^{m+n,r}\de_{j,k}\,. &
             \end{array}
\eq

We pass from the canonical basis  (\ref{inv5}), (\ref{ocb}) in $\gh_0$
to the  basis of simple coroots
\beq{pid}
 \ti\Pi^\vee=\{H_{\ti\al_j}\equiv
 H_j=\frac{2\ti\al_j}{(\ti\al_j,\ti\al_j)}\,,~\ti\al_j\in\ti\Pi\,,~(\ref{pi})\}
\eq


Let $\Xi$  be the system of fundamental weights of $\gg$
\beq{fw}
\Xi=\{\varpi_1,\ldots,\varpi_l\,|\,
\lan\varpi_j,\al_k\ran=\de_{jk}\,,~\al_k\in\Pi\}\,.
\eq
 Then the fundamental co-weights $\ti\Xi$ of the invariant subalgebra $\gg_0$ are defined as
 \beq{ifw}
 \ti\varpi^\vee_j=\left\{\begin{array}{lc}
  \varpi^\vee_j\,, & {\rm if~}\nu(\al_j)=\al_j\,, \\
                    \sum_{m=0}^{r-1}\nu^m(\varpi^\vee_j) &  {\rm if~}\nu(\al_j)\neq\al_j
                  \end{array}
                  \right.
\eq
 The system $\ti\Psi^\vee=\{ \ti\varpi^\vee_j \}$ is dual to the system of
 simple roots $\ti\Pi$ of $\gg_0$
 ($\lan\ti\varpi^\vee_j,\ti\al_k\ran=\de_{jk}$).


Let $ \ti P^\vee$ be the coweight lattice of $\gg_0$ generated by
 the fundamental coweights
 \beq{cwl1}
 \ti P^\vee=\sum_{j=1}^nm_j\ti\varpi^\vee_j\,,~m_j\in\mZ\,,
 \eq
 and $ \ti Q^\vee\subseteq \ti P^\vee$ is the  coroot lattice
 \beq{cwl2}
 \ti Q^\vee=\sum_{j=1}^nm_j\ti\al^\vee_j\,,~m_j\in\mZ\,.
 \eq
It follows from (\ref{inr}) and (\ref{ifw}   ) that both lattices are $\nu$-invariant
\beq{sin}
\nu(\ti P^\vee)=\ti P^\vee\,,~\nu(\ti Q^\vee)=\ti Q^\vee\,.
 \eq
Moreover, $\ti P^\vee\subset P$, while $\ti Q^\vee\nsubseteq Q$.


Let
\beq{iwc}
 \ti C^+=\{\,x\in\gh^\mR_0\,|\,\lan\ti\al,x\ran> 0\}\,.
 \eq
 be the positive Weyl chamber
and $ W_0$ is the Weyl group generated by the reflections of the root system $\ti R$.
Then any regular element of $\Re e\, \gh_0$ belongs to the $W_0$-orbit of $\bfu\in\ti C^+$.

By means of $\ti Q^\vee$ and $ W_0$ define  affine Weyl group $\ti W_a$
\beq{q250}
\ti W_a=W_0\ltimes\ti Q^\vee
 \eq
acting on $\Re e\, \gh_0$ as
\beq{sh}
x\to x-\lan\ti\al,x\ran\ti\al^\vee+k\ti\be^\vee\,,~~~\ti\al^\vee\,,
\ti\be^\vee\in\ti R^\vee~~k\in\mZ\,.
\eq
 The Weyl alcoves are
connected components of the set $\gh^\mR_0\setminus\{\lan\ti\al,x\ran\in\mZ\}$.
 Their closure are fundamental domains of the $\ti W_a$-action. The group $\ti W_a$ acts
 transitively on the set of the Weyl alcoves.


By means $\ti P^\vee$ define the
  semidirect product
 \beq{q25}
\ti W'_a=W_0\ltimes \ti P^\vee\supseteq \ti W_a\,.
 \eq
 The connected components of the quotient $\Re e\,\gh_0/\ti W'_a$ are alcoves
 \beq{wap}
 \ti C'_{alc}=\Re e\,\gh_0/\ti W'_a\,.
\eq

 The quotient group $\ti\cZ=\ti P^\vee/\ti Q^\vee\sim\ti W'_a/\ti W_a$. These groups
 have the structure
 \beq{tce}
 \ti\cZ(G)=
 \left\{
 \begin{array}{ll}
   \mZ_2 & \gg=A_{2n+1}\,,\,D_{n+1}\,, \\
   Id & \gg= A_{2n}\,,\,D_{4}\,,\,E_6\,.
 \end{array}
 \right.
 \eq



\begin{thebibliography}{99}

\small{


\bibitem{Atiyah}
M. Atiyah, \emph{Vector bundles over an elliptic curve,} Proc.
London Math.  Soc.  {\bf 7} (1957) 414--452.


\bibitem{B}\
Bateman, Harry,
\emph{ Higher Transcendental Functions}, Volume II,
 (1953).

\bibitem{BS}
I.N. Bernstein, O.V. Schwarzman, \emph{ Chevalley's theorem for
complex crystallographic Coxeter groups,} Functional Analysis and
Its Applications  {\bf 12} (4) (1978)  308--310


\bibitem{BST} A.J. Bordner, R. Sasaki, K. Takasaki,
\emph{Calogero-Moser Models II: Symmetries and Foldings,}
arXiv:hep-th/9809068v3\,,\\
 A. J. Bordner, E. Corrigan and R. Sasaki,
\emph{Generalised Calogero-Moser Models and Universal Lax Pair Operators},
Prog. Theor. Phys. (1999) 102 (3): 499-529.  arXiv:hep-th/9905011\,.

\bibitem{BDOZ}
H.W. Braden, V.A. Dolgushev, M.A. Olshanetsky, A.V. Zotov,
\emph{Classical R-Matrices and the Feigin-Odesskii Algebra via Hamiltonian and Poisson Reductions},
J.Phys. {\bf A36}, (2003), 6979-7000, arXiv:hep-th/0301121\,.

\bibitem{CLOZ}
Yu.Chernyakov, A.M.Levin, M.Olshanetsky, A.Zotov.
\emph{Quadratic algebras related to elliptic curves},
Theoretical and Mathematical Physics, {\bf 156} (2008),  1103-1122\,,  arXiv:0710.1072.

\bibitem{ER}
B. Enriques and V. Rubtsov, {\em Hitchin systems, higher Gaudin
operators and r-matrices}, Math. Res. Lett. 3 (1996) 343--357.

\bibitem{ES} P. Etingof, O. Schiffmann,
\emph{Twisted traces of quantum intertwiners
and quantum dynamical R-matrices corresponding to generalized Belavin-Drinfeld triples},
arXiv:math/0003109\,.

\bibitem{ES1} P. Etingof, O. Schiffmann,
\emph{ On the moduli space of classical dynamical r-matrices},
 arXiv: math/0005282. – 2000.

\bibitem{FP} L. Feher, B.G. Pusztai, {\em Generalizations of Felder's elliptic
dynamical r-matrices associated with twisted loop algebras of
self-dual Lie algebras},     Nucl. Phys. B621 (2002) 622--642;
arXiv:math/0109132.

\bibitem{FP1} L. Feher, B.G. Pusztai,
\emph{ Spin Calogero models obtained from dynamical r-matrices and geodesic motion},

\bibitem{FP2} L. Feher, B.G. Pusztai,
\emph{Twisted spin Sutherland models from quantum Hamiltonian reduction,}
Journal of Physics A: Mathematical and Theoretical, {\bf 41}, (2008),  194009.

Nuclear Physics B. {\bf 734} (2006),  304-325.

\bibitem{Fe}
G.Felder, \emph{The KZB equations on Riemann surfaces}, arXiv:hep-th/9609153.

\bibitem{FGK}
G.Felder, K.Gawedzki, A.Kupiainen, \emph{Spectra Of Wess-Zumino-Witten Models With Arbitrary Simple Groups}, Commun. Math.
Phys. {117} (1988) 127-158.

\bibitem{FW}
G.Felder, Ch.Wieczerkowski, \emph{Conformal blocks on elliptic curves and the Knizhnik--Zamolodchikov--Bernard equations},
Commun. Math. Phys. { 176} (1996), 133-162; arXiv:hep-th/9411004.

\bibitem{GH}
J.Gibbons, and T.Hermsen,
 \emph{A generalization of the Calogero-Moser systems,} Physica \textbf{11D}
(1984), 337-348.

\bibitem{GN}
A.  Gorsky, N.  Nekrasov, {\em  Elliptic Calogero-Moser system
 from two dimensional current algebra},
arXiv:hep-th/9401021\,.


\bibitem{Hi}\ Hitchin N.,
{\em Stable  bundles and Integrable Systems},
Duke Math. Journ., {\bf 54}, (1987) 91-114\,.

\bibitem{DHP} E. D'Hoker, D.H. Phong,\\
\emph{Calogero-Moser Lax Pairs with Spectral Parameter for General Lie Algebras},
Nucl.Phys. {\bf B530}, (1998) 537-610, arXiv:hep-th/9804124\,,\\
\emph{Calogero-Moser and Toda Systems for Twisted and Untwisted Affine Lie Algebras},
Nucl.Phys. {\bf B530}, (1998) 611-640, arXiv:hep-th/9804125\,.

\bibitem{HM}
J.C. Hurtubise, E. Markman,\emph{ Calogero–Moser systems and Hitchin systems },
Comm. Math. Phys., {\bf 223} (2001), 533-552.

\bibitem{LX}
Luen-Chau Li, Ping Xu, \emph{Integrable spin Calogero-Moser systems}
 Commun.Math.Phys., {\bf 231} (2002), 257-286.

 \bibitem{Ka}
V.~Kac,\emph{ Automorphisms of finite order of semisimle Lie algebras},
Funct.Anal. Applic. {\bf 3} (1969) 94-96.

  \bibitem{Ka1}
V.~Kac,\emph{ Infinite dimensional Lie algebras}, Cambridge Univ. Press\,, 1990\,.

 \bibitem{KT}
S. P. Kumar, J. Troost, \emph{Geometric construction of elliptic integrable systems and
N= 1* superpotentials},
JHEP, {\bf 0201} (2002) 020, hep-th/0112109.

 \bibitem{KuT}
G.Kuroki,  and T.Takebe,\emph{ Twisted Wess-Zumino-Witten models on elliptic curves},
 Communications in mathematical physics, {\bf 190} (1997), 1-56.

\bibitem{LOZ1} A.M. Levin, M.A. Olshanetsky, A. Zotov,
\emph{Hitchin Systems - Symplectic Hecke Correspondence and Two-dimensional Version,}
Communications in Mathematical Physics,
 {\bf 236} (2003) 93-133, arXiv:nlin/0110045\,.

\bibitem{LOSZ1} A.M. Levin, M.A. Olshanetsky, A.V. Smirnov, A.V. Zotov,
{\em Characteristic Classes and Hitchin Systems. General Construction}, Commun. Math. Phys.,
 Vol. 316, Num. 1 (2012) 1-44,
arXiv:1006.0702 [math-ph].

\bibitem{LOSZ2} A.M. Levin, M.A. Olshanetsky, A.V. Smirnov, A.V. Zotov,
{\em Calogero–Moser systems for simple Lie groups and characteristic classes of bundles},
Journal of Geometry and Physics 62
(2012) 1810-1850, arXiv:1007.4127.

\bibitem{LOSZ3} A.M. Levin, M.A. Olshanetsky, A.V. Smirnov, A.V. Zotov,
 \emph{Hecke transformations of conformal blocks in WZW theory. I.
 KZB equations for non-trivial bundles},
  SIGMA, {\bf 8} (2012) ¹. 095., arXiv:1207.4386.

\bibitem{Lo}
E. Looijenga, \emph{Root systems and elliptic curves,} Invent. Math.
38 (1976), 17--32.


\bibitem{NS}
 M.S. Narasimhan and C.S. Seshadri,
  \emph{Stable and unitary vector bundles on a compact Riemann surface,}
   Ann. of Math. { 82} (1965) 540--64.

\bibitem{N}
   N.  Nekrasov, \emph{Holomorphic Bundles and Many-Body Systems},
 Commun.  Math.  Phys.  { 180} (1996) 587--604; hep-th/9503157\,.


\bibitem{PR}
G. Pappas,  M. Rapoport,
\emph{ Twisted loop
groups and their affine flag varieties,} Advances in Mathematics, {\bf 219}, (2008): 118-198.

\bibitem{PS}
 A.Presley,  G.Segal, \emph{Loop Groups}, Oxford: Clarendon Press, 1986.


 \bibitem{VO}
 E.Vinberg and A.Onishik, \emph{Seminar on Lie groups and algebraic groups}, (in Rus-
sian), Moscow Nauka 1988

\bibitem{We}\
Weil, A. \emph{Elliptic functions according to Eisenstein and Kronecker}, (Vol. 88).(1976)
 Berlin-Heidelberg-New York: Springer.

 \bibitem{Wo}
S.Wojciechowski,
 \emph{An integrable marriage of the Euler equations with the Calogero-Moser
systems,} Phys. Lett. A, \textbf{111} (1985), 101-103.

}

\end{thebibliography}
\end{document}